\documentclass[aps, prd, floatfix, nofootinbib, superscriptaddress, twocolumn]{revtex4-2}

\usepackage{latexsym}
\usepackage{amsmath}
\usepackage{amssymb}
\usepackage{amsfonts}
\usepackage{textcomp}
\usepackage{color}
\usepackage{CJKutf8}

\usepackage[mathscr,scaled=1.15]{urwchancal}
\DeclareFontFamily{OT1}{pzc}{}
\DeclareFontShape{OT1}{pzc}{m}{it}%
{<-> s * [1.15] pzcmi7t}{}
\DeclareMathAlphabet{\mathpzc}{OT1}{pzc}{m}{it}

\usepackage{color}

\usepackage{supertabular}
\usepackage{placeins}
\usepackage{epsfig}
\usepackage{graphicx}

\definecolor{purple}{rgb}{0.5,0,0.5}
\definecolor{blue}{rgb}{0.0,0,0.9}
\definecolor{prdblue}{rgb}{0.133,0.118,0.498}
\usepackage[colorlinks=true, pdfstartview=FitV, linkcolor=prdblue, citecolor= prdblue, urlcolor=prdblue]{hyperref}

\makeatletter

\setbox0\hbox{$\xdef\scriptratio{\strip@pt\dimexpr
    \numexpr(\sf@size*65536)/\f@size sp}$}

\newcommand{\scriptveryshortarrow}[1][3pt]{{%
    \hbox{\rule[\scriptratio\dimexpr\fontdimen22\textfont2-.2pt\relax]
               {\scriptratio\dimexpr#1\relax}{\scriptratio\dimexpr.4pt\relax}}%
   \mkern-4mu\hbox{\let\f@size\sf@size\usefont{U}{lasy}{m}{n}\symbol{41}}}}

\makeatother

\begin{document}

\begin{CJK}{UTF8}{song}

\title{$\,$\\[-6ex]\hspace*{\fill}{\normalsize{\sf\emph{Preprint no}.\ NJU-INP 070/23}}\\[1ex]%
Empirical Determination of the Pion Mass Distribution}

\date{2023 March 15}

\author{Y-Z.~Xu
       $^{\href{https://orcid.org/0000-0003-1623-3004}{\textcolor[rgb]{0.00,1.00,0.00}{\sf ID}},}$}
\affiliation{Dpto.~Ciencias Integradas, Centro de Estudios Avanzados en Fis., Mat. y Comp., Fac.~Ciencias Experimentales, Universidad de Huelva, Huelva 21071, Spain}
\affiliation{Dpto. Sistemas F\'isicos, Qu\'imicos y Naturales, Univ.\ Pablo de Olavide, E-41013 Sevilla, Spain}

\author{K.~Raya%
       $^{\href{https://orcid.org/0000-0001-8225-5821}{\textcolor[rgb]{0.00,1.00,0.00}{\sf ID}},}$}
\affiliation{Dpto.~Ciencias Integradas, Centro de Estudios Avanzados en Fis., Mat. y Comp., Fac.~Ciencias Experimentales, Universidad de Huelva, Huelva 21071, Spain}

\author{Z.-F.~Cui
       $^{\href{https://orcid.org/0000-0003-3890-0242}{\textcolor[rgb]{0.00,1.00,0.00}{\sf ID}},}$}
\affiliation{School of Physics, Nanjing University, Nanjing, Jiangsu 210093, China}
\affiliation{Institute for Nonperturbative Physics, Nanjing University, Nanjing, Jiangsu 210093, China}

\author{C.\,D.~Roberts%
       $^{\href{https://orcid.org/0000-0002-2937-1361}{\textcolor[rgb]{0.00,1.00,0.00}{\sf ID}},}$}
\affiliation{School of Physics, Nanjing University, Nanjing, Jiangsu 210093, China}
\affiliation{Institute for Nonperturbative Physics, Nanjing University, Nanjing, Jiangsu 210093, China}

\author{J.~Rodr\'iguez-Quintero%
       $^{\href{https://orcid.org/0000-0002-1651-5717}{\textcolor[rgb]{0.00,1.00,0.00}{\sf ID}},}$}
\affiliation{Dpto.~Ciencias Integradas, Centro de Estudios Avanzados en Fis., Mat. y Comp., Fac.~Ciencias Experimentales, Universidad de Huelva, Huelva 21071, Spain}

\begin{abstract}
\vspace*{-3ex}

\begin{small}
\noindent\href{mailto:phycui@nju.edu.cn}{phycui@nju.edu.cn} (ZFC);
\href{mailto:cdroberts@nju.edu.cn}{cdroberts@nju.edu.cn} (CDR);
\href{mailto:ose.rodriguez@dfaie.uhu.es}{jose.rodriguez@dfaie.uhu.es} (JRQ)
\end{small}
\\

Existing pion+nucleus Drell-Yan and electron+pion scattering data are used to develop ensembles of model-independent representations of the pion generalised parton distribution (GPD).  Therewith, one arrives at a data-driven prediction for the pion mass distribution form factor, $\theta_2$.  Compared with the pion elastic electromagnetic form factor, $\theta_2$ is harder: the ratio of the radii derived from these two form factors is $r_\pi^{\theta_2}/r_\pi = 0.79(3)$.  Our data-driven predictions for the pion GPD, related form factors and distributions should serve as valuable constraints on theories of pion structure.
\end{abstract}

\maketitle

\end{CJK}


\noindent\emph{1.$\;$Introduction} ---
Science is operating, building, and planning high-luminosity, high-energy facilities in order to reveal the source of the mass of visible material in the Universe \cite{Adams:2018pwt, Quintans:2022utc, Aguilar:2019teb, Brodsky:2020vco, Chen:2020ijn, Anderle:2021wcy, Arrington:2021biu, AbdulKhalek:2021gbh, Wang:2022xad, Carman:2023zke}.  That matter is chiefly built from the nuclei that can be found on Earth, which are themselves composed of neutrons and protons (nucleons) bound together by the exchange of $\pi$-mesons (pions) \cite{Machleidt:2011zz} -- and other contributors at shorter ranges.  Partly, therefore, some of the mass is generated by Higgs boson couplings to matter fields in the Standard Model (SM) \cite{Aad:2012tfa, Chatrchyan:2012xdj,  Englert:2014zpa, Higgs:2014aqa}.

However, insofar as nucleons and pions are concerned, the Higgs-generated mass component is only a small part.
Regarding the nucleon -- a bound state with mass $m_N \approx 940\,$MeV, yet built from three light valence quarks, $\sum m_q \approx 9\,$MeV, Higgs boson couplings are directly responsible for $\lesssim 1$\% of its mass: the remainder has its origin in some other mechanism.
The story for the pions, a triplet of three electric charge states, whose positive member is a bound state of a valence $u$-quark and a valence $\bar d$ quark, is very subtle because pions are the SM's (would-be) Nambu-Goldstone bosons whose emergence can be traced to exactly the same source as 99\% of the nucleon mass, whatever that may be.
This second source is called emergent hadron mass (EHM) \cite{Roberts:2016vyn, Roberts:2020udq, Roberts:2020hiw, Krein:2020yor, Roberts:2021xnz, Roberts:2021nhw, Binosi:2022djx, Papavassiliou:2022wrb, Ding:2022ows, Roberts:2022rxm, Ferreira:2023fva}.

Modern theory indicates \cite{Roberts:2016vyn, Roberts:2020udq, Roberts:2020hiw, Krein:2020yor, Roberts:2021xnz, Roberts:2021nhw, Binosi:2022djx, Papavassiliou:2022wrb, Ding:2022ows, Roberts:2022rxm, Ferreira:2023fva} that EHM is a feature of SM strong interactions, \emph{i.e}., quantum chromodynamics (QCD).  Further, that owing to its Nambu-Goldstone boson character, expressed in QCD symmetry identities \cite{Bhagwat:2007ha, Brodsky:2010xf, Qin:2014vya}, pion properties provide the clearest window onto EHM \cite{Roberts:2016vyn, Roberts:2020udq, Roberts:2020hiw, Roberts:2021xnz, Roberts:2021nhw}.
It is thus imperative that theory provide constraints on the distribution of mass within the pion in advance of new facility operation, so forthcoming experiments can truly test the EHM paradigm, avoiding a common pattern of models being developed to describe novel data after they have been collected.
To this end, herein, we use existing data on pion valence quark distribution functions (DFs) \cite{Corden:1980xf, Badier:1983mj, Betev:1985pg, Conway:1989fs} and the pion elastic electromagnetic form factor \cite{Amendolia:1984nz, Amendolia:1986wj, Volmer:2000ek, Horn:2006tm, Tadevosyan:2007yd, Blok:2008jy, Huber:2008id} to develop data-based predictions for the three-dimensional (3D) structure of the pion and, therefrom, the pion mass distribution.

\medskip

\noindent\emph{2.$\;$Pion Valence-quark Distribution Function and the Hadron Scale} ---
Consider the pion valence $u$ quark DF, ${\mathpzc u}^\pi(x;\zeta)$, which is the probability density for finding a valence $u$-quark with light-front momentum fraction $x$ of the pion's total momentum when the pion is probed at energy scale $\zeta$ \cite{Holt:2010vj}.  In the ${\cal G}$-parity symmetry limit, which is an accurate reflection of Nature, $\bar {\mathpzc d}^\pi(x;\zeta)={\mathpzc u}^\pi(x;\zeta)$.

Extant data relevant to extraction of ${\mathpzc u}^\pi(x;\zeta)$ has been obtained using the Drell-Yan process \cite{Peng:2016ebs, SeaQuest:2021zxb} $\pi + A \to \ell^+ \ell^- + X$, where $\ell$ is a lepton, $A$ is a nuclear target, and $X$ denotes the debris produced by the deeply inelastic reaction \cite{Corden:1980xf, Badier:1983mj, Betev:1985pg, Conway:1989fs}.  New data will be obtained using Drell-Yan \cite{Adams:2018pwt, Quintans:2022utc} and other processes \cite{Aguilar:2019teb, Brodsky:2020vco, Chen:2020ijn, Roberts:2021nhw, Anderle:2021wcy, Arrington:2021biu}.

The most malleable set of Drell-Yan data \cite[E615]{Conway:1989fs} was collected at a large resolving scale, \emph{viz}.\ $\zeta_5=5.2\,$GeV.  In this case, QCD perturbation theory can be employed in the analysis and a reasonable interpretative basis is provided by the quark and gluon parton degrees-of-freedom used to define the QCD Lagrangian density.

On the other hand, development of a theory prediction for ${\mathpzc u}^\pi(x;\zeta)$ is best begun at the hadron scale, $\zeta_{\cal H}\ll \zeta_5$, whereat dressed-quark and -antiquark quasiparticle degrees-of-freedom can be used to deliver symmetry-preserving, parameter-free predictions for pion properties \cite{Gao:2017mmp, Chen:2018rwz, Ding:2019qlr, Ding:2019lwe, Cui:2020tdf}.  Of particular importance and utility is the fact that the bound-state's quasiparticle degrees-of-freedom carry all measurable properties of the hadron at this scale, including the light-front momentum; hence \cite{Ding:2019qlr, Ding:2019lwe}:
\begin{equation}
\label{EqSymmetric}
{\mathpzc u}^\pi(x;\zeta_{\cal H}) = {\mathpzc u}^\pi(1-x;\zeta_{\cal H})\,.
\end{equation}

In QCD perturbation theory, the evolution of ${\mathpzc u}^\pi(x;\zeta)$ with changing scale, $\zeta$, is described by the DGLAP equations \cite{Dokshitzer:1977sg, Gribov:1971zn, Lipatov:1974qm, Altarelli:1977zs}.  A nonperturbative extension is explained in Refs.\,\cite{Cui:2019dwv, Cui:2021mom, Cui:2022bxn}.  It is based on the following proposition.\\
\hspace*{0.75em}\parbox[t]{0.95\linewidth}{{\sf P1} -- \emph{There exists at least one effective charge, $\alpha_{1\ell}(k^2)$, such that, when used to integrate the leading-order perturbative DGLAP equations, an evolution scheme for parton DFs is defined that is all-orders exact}.}  
\smallskip

\noindent Such charges are discussed elsewhere \cite{Grunberg:1980ja, Grunberg:1982fw, Dokshitzer:1998nz}.  They need not be process-independent (PI); hence, not unique.
Nevertheless, an efficacious PI charge is not excluded; and that discussed in Refs.\,\cite{Cui:2019dwv, Cui:2020tdf, Raya:2021zrz} has proved suitable.
Connections with experiment and other nonperturbative extensions of QCD's running coupling are given in Refs.\,\cite{Deur:2016tte, Deur:2022msf, Deur:2023}.

Experiment and theory can now be joined at $\zeta=\zeta_{\cal H}$ because {\sf P1} and Eq.\,\eqref{EqSymmetric} entail the following evolution equation \cite[Sec.\,VII]{Raya:2021zrz}: $\forall \zeta \geq \zeta_{\cal H}$,
\begin{equation}
\label{EqxnzzH}
\langle x^n\rangle_{{\mathpzc u}_\pi}^\zeta
= \langle x^n\rangle_{{\mathpzc u}_\pi}^{\zeta_{\cal H}}
\left( \langle 2 x \rangle_{{\mathpzc u}_\pi}^\zeta\right)^{\gamma_0^n/\gamma_0^1},
\end{equation}
where the DF Mellin moments are
\begin{equation}
\label{EqMellin}
\langle x^n \rangle_{{\mathpzc u}_\pi}^{\zeta}
:= \langle x^n  {\mathpzc u}^\pi(x;\zeta)\rangle
= \int_0^1 dx\,x^n\, {\mathpzc u}^\pi(x;\zeta)\,
\end{equation}
and, with $n_f=4$ flavours of active quarks,
\begin{equation}
\label{eq:Dfn}
\gamma_0^n = -\frac 4 3 \left( 3 + \frac{2}{(n+1)(n+2)} - 4 \sum_{j=1}^{n+1} \frac 1 j  \right) \,.
\end{equation}
Owing to Eq.\,\eqref{EqSymmetric}, $\langle 2 x \rangle_{{\mathpzc u}_\pi}^{\zeta_{\cal H}} = 1$.

Using Eq.\,\eqref{EqxnzzH}, any set of valence-quark DF Mellin moments at a scale $\zeta$ can be converted into an equivalent set of hadron-scale moments; hence, one has a direct mapping between the pion valence-quark DF at any two scales.  Crucially, the leading moment, $\langle 2 x \rangle_{{\mathpzc u}_\pi}^\zeta$, is the kernel of the mapping; so, although existence of $\alpha_{1\ell}(k^2)$ is essential, its pointwise form is largely immaterial.

If one analyses the data in Ref.\,\cite[E615]{Conway:1989fs} using methods that ensure consistency with QCD endpoint $(x\simeq 0, 1)$ constraints, then realistic connections with SM properties can be drawn \cite{Cui:2021mom, Cui:2022bxn}.  Two such studies are available: Ref.\,\cite{Aicher:2010cb}, whose result we label ${\mathpzc u}^\pi_{\rm A}(x;\zeta_5)$; and the Mellin-Fourier analysis in Ref.\,\cite{Barry:2021osv}, reappraised in Ref.\,\cite{Cui:2021mom} and labelled ${\mathpzc u}^\pi_{\rm B}(x;\zeta_5)$ herein.

\smallskip

\noindent{\sf S1} -- In proceeding, we do not bind ourselves to any single fit to E615 data.  Instead, we consider a large array of possibilities with the same qualitative character.  To achieve this generalisation, we first suppose that a fair approximation to any such DF is provided by
\begin{equation}
\label{BestDF}
{\mathpzc u}^\pi(x;[\alpha_i];\zeta) = {\mathpzc n}_{\mathpzc u}^\zeta
x^{\alpha_1^\zeta} (1-x)^{\alpha_2^\zeta} (1 + \alpha_3^\zeta x^2)\,,
\end{equation}
where $n_{\mathpzc u}$ is fixed by baryon-number conservation, \emph{i.e}.,
$\langle x^0 \rangle_{{\mathpzc u}_\pi}^{\zeta}=1$.
This is a weak assumption \cite{Cui:2021mom}: any of the forms commonly used in fitting data would serve just as well.  Then, we proceed as follows.
(\emph{i}) Determine central values of $\{\alpha_i^\zeta|i=1,2,3\}$ via a least-squares fit to the ${\mathpzc u}^\pi_{\rm A(B)}(x;\zeta_5)$ data.
%
(\emph{ii}) Generate a new vector $\{\alpha_i^\zeta|i=1,2,3\}$, each element of which is distributed randomly around its best-fit value.
(\emph{iii})
Using the DF obtained therewith, evaluate
\begin{equation}
\chi^2 = \sum_{l=1}^{N}\frac{({\mathpzc u}^\pi_{\rm A(B)}(x_l;[\alpha_i];\zeta_5)-u_j)^2}{\delta_l^2}\,,
\label{eq:chi2}
\end{equation}
where $\{(x_l,u_l\pm\delta_l) | l=1,\ldots\,N=40\}$ are the measured $x$ points in the E615 data set.  This $\{\alpha_i\}$ configuration is accepted with probability
\begin{equation}
\label{EqProb}
{\mathpzc P}  = \frac{P(\chi^2;d)}{P(\chi_0^2;d)} \,, \;
P(y;d) = \frac{(1/2)^{d/2}}{\Gamma(d/2)} y^{d/2-1} {\rm e}^{-y/2}\,,
\end{equation}
where $d=N-3$ and $\chi_0^2 \approx d$ locates the maximum of the $\chi^2$-probability density, $P(\chi^2;d)$.
(\emph{iv}) Repeat (\emph{ii}) and (\emph{iii}) until one has a $K \gtrsim 100$-member ensemble of DFs in both cases A, B.
%

\smallskip

\noindent{\sf S2} -- These ensembles are representative of the $\zeta=\zeta_5$ E615 A(B) data analyses.
Using Eq.\,\eqref{EqxnzzH}, each member of an ensemble can be evolved to $\zeta=\zeta_{\cal H}$.
That is readily achieved by
calculating a large number, $M$, of Mellin moments of the DF under consideration, $\{\langle x^m\rangle_{{\mathpzc u}^\pi}^{\zeta_5} | 1\leq m \leq M\}$;
evolving each moment to $\zeta=\zeta_{\cal H}$;
and then reconstructing the equivalent hadron-scale DF in the form
\begin{equation}
\label{hadronscaleDF}
{\mathpzc u}^\pi(x;\zeta_H) = {\mathpzc n}_0 \ln ( 1+ x^2 (1-x)^2/\rho^2)
\end{equation}
by choosing $\rho$ such that a best least-squares fit is obtained to the target set of hadron-scale moments.  The function in Eq.\,\eqref{hadronscaleDF} is efficacious because
it is symmetric, as required by Eq.\,\eqref{EqSymmetric};
consistent with QCD constraints on the endpoint behaviour of valence-quark DFs \cite[Sec.\,2]{Cui:2021mom};
and flexible enough to express the dilation that EHM is known to produce in pion DFs \cite{Ding:2019qlr, Ding:2019lwe, Cui:2020tdf}.

\begin{figure}[t]
\centerline{\includegraphics[width=0.42\textwidth]{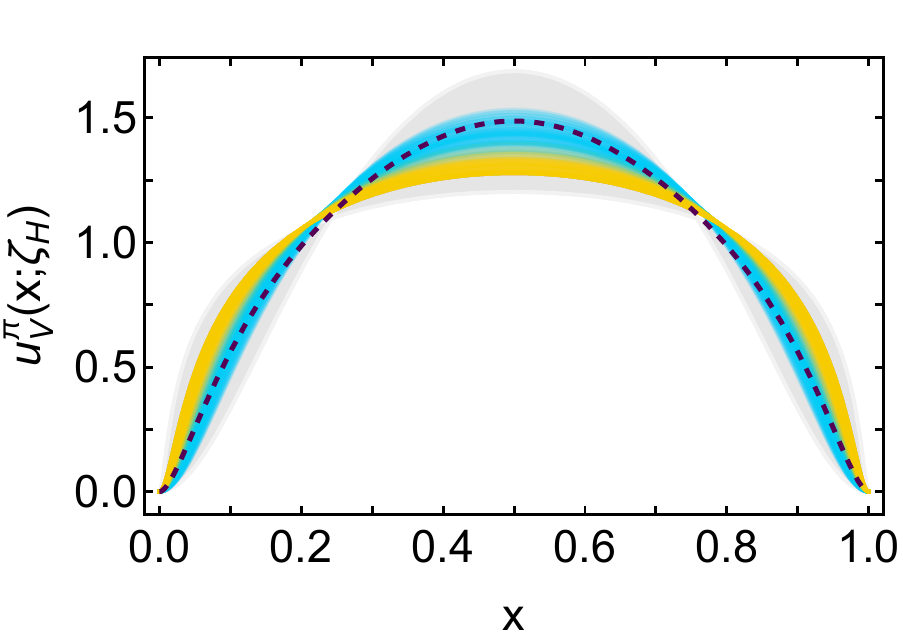}}
\caption{\label{FigE615zH}
Ensembles of $\zeta_5\to \zeta_{\cal H}$ pion valence-quark DF replicas:
${\mathpzc u}^\pi_{\rm A}(x;\zeta_{\cal H})$ \cite{Aicher:2010cb} -- blue band;
${\mathpzc u}^\pi_{\rm B}(x;\zeta_{\cal H})$ \cite[Sec.\,8]{Cui:2021mom} -- orange band.
Comparison curves:
dashed purple -- pion valence-quark DF calculated using continuum Schwinger function methods (CSMs) \cite{Cui:2020tdf};
grey band -- ensemble of valence-quark DFs developed in Ref.\,\cite{Cui:2022bxn} from results obtained using lattice Schwinger function methods \cite{Joo:2019bzr, Sufian:2019bol, Alexandrou:2021mmi}.
}
\end{figure}

\smallskip

Following {\sf S1}, {\sf S2}, one arrives at the $\zeta=\zeta_{\cal H}$ representation ensembles of E615 data drawn in Fig.\,\ref{FigE615zH}.  Evidently, the data-driven results are a fair match with modern theory predictions, albeit those based on the analysis in Ref.\,\cite{Aicher:2010cb} deliver better agreement.

\medskip

\noindent\emph{3.$\;$Reconstructing the GPD} ---
Generalised parton distributions (GPDs) are discussed in Refs.\,\cite{Belitsky:2005qn, Mezrag:2022pqk, Mezrag:2023}.  They provide an extension of one-dimensional (1D -- light-front longitudinal) DFs into 3D images of hadrons because they also express information about the distribution of partons in the plane perpendicular to the bound-state's total momentum, \emph{i.e}., within the light front.  Data that may be interpreted in terms of GPDs can be obtained via deeply virtual Compton scattering on a target hadron, $T$, \emph{viz}.\ $\gamma^\ast(q) T(p) \to \gamma^\ast(q^\prime)T(p^\prime)$, so long as at least one of the photons [$\gamma^\ast(q)$, $\gamma^\ast(q^\prime)$] possesses large virtuality, and in the analogous process of deeply virtual meson production: $\gamma^\ast(q) T(p) \to M(q^\prime)T(p^\prime)$, where $M$ is a meson.  Moreover, GPDs connect DFs with hadron form factors because any DF may be recovered as a forward limit $(p^\prime = p)$ of the relevant GPD and any elastic form factor can be expressed via a GPD-based sum rule.

Exploiting this last feature, the valence-quark DFs obtained above can be used to develop a 3D image of the pion by working with the light-front wave function (LFWF) overlap representation of GPDs.  Capitalising on the property of LFWF factorisation, which is an excellent approximation for the pion \cite[Sec.\,3]{Roberts:2021nhw}, \cite{Xu:2018eii, Zhang:2021mtn, Raya:2021zrz}, one may write the $|x|\geq|\xi|$ pion GPD as follows:
\begin{align}
H_\pi^u(x,\xi,&-\Delta^2;\zeta_H) =  \nonumber \\
& \theta(x_{-} ) \sqrt{u^\pi(x_-;\zeta_H) u^\pi(x_+;\zeta_H)} \, \Phi^\pi(z^2;\zeta_H) \,,
\label{eq:GPD}
\end{align}
where $P=(p+p^\prime)/2$;
$\Delta=p^\prime-p$;
$\xi = - [n \cdot \Delta] / [2 n \cdot P]$, with $n$ a light-like four-vector, $n^2=0$;
$x_\pm =  (x \pm \xi)/(1 \pm \xi)$;
and, with $m_\pi$ being the pion mass,
\textcolor[rgb]{0.5,0,0.5}{\begin{subequations}
\label{Kinematics}
\begin{align}
%
z^2& =\Delta_\perp^2 (1-x)^2/(1-\xi^2)^2\,, \\
\Delta_\perp^2 & = \Delta^2 (1-\xi^2) + 4 \xi^2 m_\pi^2 \,.
\end{align}
\end{subequations}}

In principle, the shape of $\Phi^\pi(z;\zeta_H)$ is determined by the pion LFWF -- see Ref.\,\cite[Eq.\,(18.b)]{Raya:2021zrz}.  However, in our data-driven approach, we exploit the following GPD sum rule, which relates the zeroth moment of the GPD to the pion elastic electromagnetic form factor:
\begin{subequations}
\label{eq:sumrule}
\begin{align}
F_\pi(\Delta^2) & \equiv F_\pi^u(\Delta^2) = \int_{-1}^1 dx \, H_\pi^u(x,0,-\Delta^2;\zeta_H)  \\
&= \int_0^1 dx \, u^\pi(x;\zeta_H) \, \Phi^\pi(\Delta^2 [1-x]^2;\zeta_H) \,,
\end{align}
\end{subequations}
where we have used Eqs.\,\eqref{Kinematics}.  Charge and baryon number conservation entail $\Phi^\pi(0;\zeta_H)\equiv 1$.

At this point, one can use available precision data on the pion elastic form factor \cite{Volmer:2000ek, Horn:2006tm, Tadevosyan:2007yd, Blok:2008jy, Huber:2008id} and a recent objective determination of the pion charge radius \cite{Cui:2021aee} to complete a reconstruction of $H_\pi^u(x,0,-\Delta^2;\zeta_H)$.  To achieve this, we proceed as follows.

\smallskip

\noindent{\sf S3} -- Neglecting logarithmic corrections, $F_\pi(\Delta^2) \propto 1/\Delta^2 $ on $\Delta^2 \gg m_N^2$ \cite{Lepage:1979za, Efremov:1979qk, Lepage:1980fj}; so, we use a $[1,2]$ Pad\'e approximant to complete Eq.\,\eqref{eq:GPD} ($y=z^2$):
\begin{equation}
\label{eq:PhiPade}
\Phi^\pi(y;\zeta_H)  = \frac{1+\lambda y}{1+\beta y + \gamma^2 y^2}\,,\;
\lambda  = \beta - \frac{r_\pi^2}{6 \langle x^2\rangle_{{\mathpzc u}_\pi}^{\zeta_{\cal H}}} \;,
\end{equation}
where the last identity follows from the standard definition of the charge radius, \emph{viz}.\
$r_\pi^2 = -6 (d/d\Delta^2)\ln F_\pi(\Delta^2)$.

Now, for every member of the ensembles obtained via {\sf S1}, {\sf S2}, drawn in Fig.\,\ref{FigE615zH}, we use Eqs.\,\eqref{eq:sumrule}, \eqref{eq:PhiPade} to obtain a best fit to available $F_\pi(\Delta^2)$ data \cite{Volmer:2000ek, Horn:2006tm, Tadevosyan:2007yd, Blok:2008jy, Huber:2008id} -- specified as $\{(\Delta^2_l,F_l \pm \delta_l) \,|\, l=1,\dots L \}$, $L=9$ -- by minimisation of a $\chi^2$ defined by analogy with Eq.\,\eqref{eq:chi2}.
In this procedure, $\beta$, $\gamma$ are fitting parameters and $\lambda$ is obtained using Eq.\,\eqref{eq:PhiPade} with $r_\pi = 0.64\,$fm \cite{Cui:2021aee}.
We thereby obtain $100_{\rm A}+100_{\rm B}$ pairs $(\beta,\gamma)$.

\smallskip

\noindent{\sf S4} -- Finally, we introduce a spread in the $r_\pi$ values, generated using a Gaussian distribution centred on $r_\pi=0.64\,$fm with width $\delta_{r_\pi}=0.02\,$fm.  This width is twice the uncertainty calculated in Ref.\,\cite{Cui:2021aee}; hence, we thereby generate a conservatively constrained $r_\pi$ distribution.

\begin{table}[t]
    \caption{\label{tab:PhiPade}
   Data-constrained GPD characterisation parameters, Eqs.\,\eqref{hadronscaleDF}, \eqref{eq:PhiPade}, obtained using {\sf S4}.
    Regarding the lQCD row, derived from the valence-quark DF ensemble in Ref.\,\cite{Cui:2022bxn}: 
    given the range of $\rho$ variation in the ensemble, the random selection of replicas was made from $\log{\rho} \in (-3,3)$, with the corresponding standard deviation translated into an asymmetric uncertainty.
    (Each DF ensemble has $K\gtrsim 100$ elements: A derived from Ref.\,\cite{Aicher:2010cb} and B from \cite[Sec.\,8]{Cui:2021mom}.  Values of $r_\pi$ are displayed and confirm that $r_\pi$ was generated with a Gaussian probability centered at $0.64\,$fm \cite{Cui:2021aee} with width $0.02\,$fm.)
    }
    \centering
    \begin{tabular}{l|l|lll}
            &  $r_\pi/{\rm fm}$ & \; $\bar{\rho}(\delta_\rho)$ & $\bar{\beta}(\delta_\beta)\times$GeV & $\bar{\gamma}(\delta_\gamma)\times$GeV\\ \hline
           A & 0.640(20) & \; 0.060(16) & 6.30(51) & 5.20(50)  \\
           B & 0.638(18) & \; 0.025(7) & 6.14(46) & 6.15(60)   \\
           lQCD & 0.639(19) & \; $0.041_{-0.023}^{+0.054}$ & 6.34(52) & 6.10(91)
    \end{tabular}
\end{table}

In concert with this $r_\pi$ distribution, the values of $(\beta,\gamma)$ are now varied at random about the best-fit values obtained in {\sf S3}; and one selects a data-driven GPD, $H$, using Eqs.\,\eqref{eq:GPD}\,--\,\eqref{eq:PhiPade}, with probability
${\cal P}_H = {\cal P}_{H | {\mathpzc u}} \, {\cal P}_{\mathpzc u}$,
where both the conditional probability of the GPD given a particular DF, ${\cal P}_{H | {\mathpzc u}}$, and that of the DF, $ {\cal P}_{\mathpzc u}$, are given by Eqs.\,\eqref{EqProb} evaluated with their respective $\chi^2$ functions, defined from DF and elastic form factor data.

\begin{figure}[t]
\vspace*{6ex}

\leftline{\hspace*{0.5em}{\large{\textsf{A}}}}
\vspace*{-7ex}
\hspace*{1em}\includegraphics[width=0.41\textwidth]{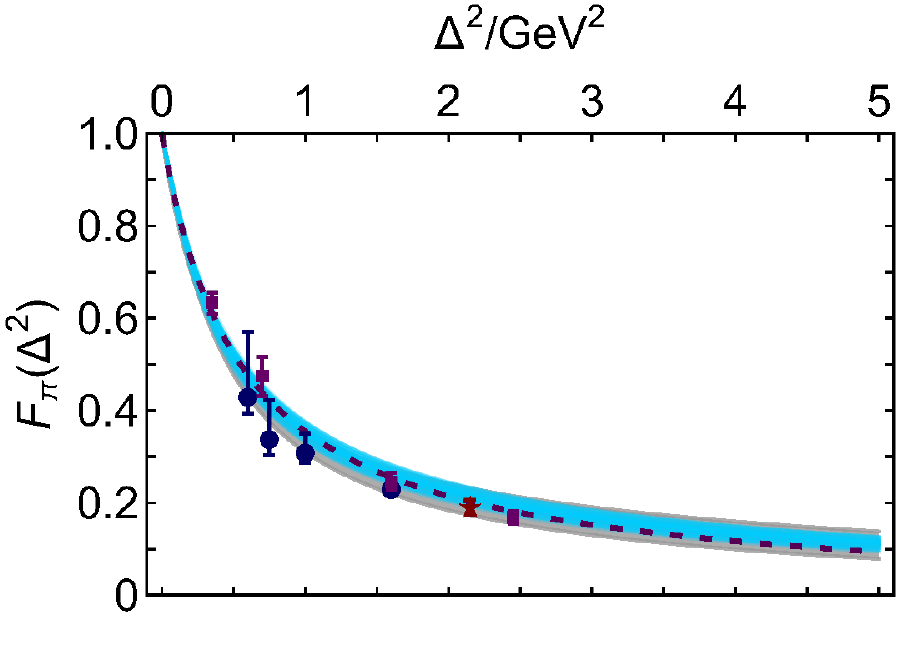}
\vspace*{-4ex}

\leftline{\hspace*{0.5em}{\large{\textsf{B}}}}
\vspace*{-3ex}
\hspace*{1em}\includegraphics[width=0.41\textwidth]{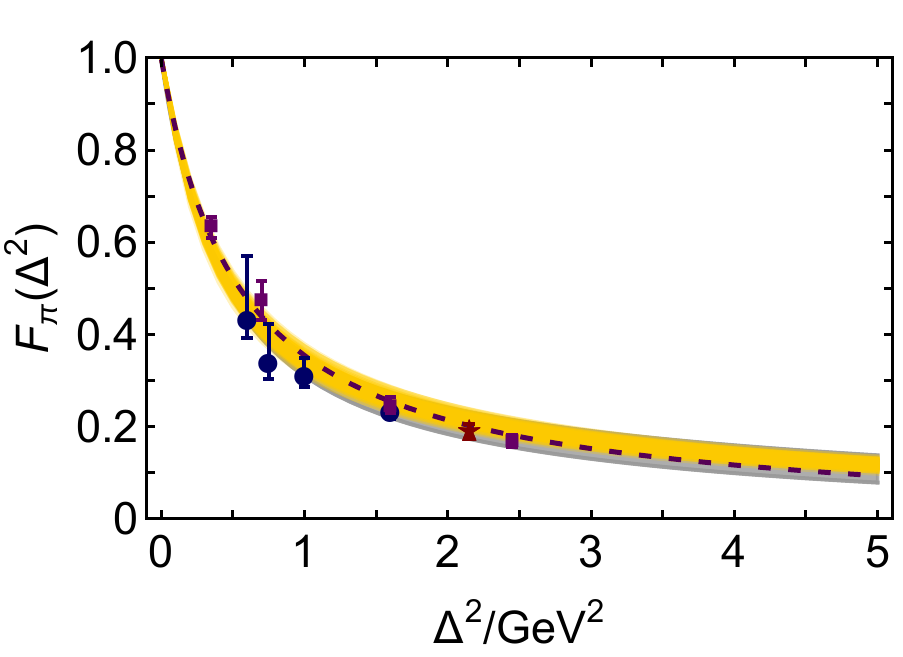}

\caption{\label{FigFpi}
Pion elastic electromagnetic form factor, $F_\pi(\Delta^2)$, obtained from Eq.\,\eqref{eq:sumrule} using the GPD ensembles generated via {\sf S4}.
\emph{Panel A}. DFs ${\mathpzc u}^\pi_{\rm A}(x;\zeta_{\cal H})$ \cite{Aicher:2010cb} (blue band).
\emph{Panel B}. DFs ${\mathpzc u}^\pi_{\rm B}(x;\zeta_{\cal H})$ \cite[Sec.\,8]{Cui:2021mom} (orange band).
Comparison curves:
dashed purple  -- $F_\pi(\Delta^2)$ calculated using CSMs \cite[Sec.\,4B]{Roberts:2021nhw}, \cite{Chen:2018rwz};
grey band -- $F_\pi(\Delta^2)$ ensemble obtained with valence-quark DFs developed in Ref.\,\cite{Cui:2022bxn} from results obtained using lattice Schwinger function methods \cite{Joo:2019bzr, Sufian:2019bol, Alexandrou:2021mmi}.
The form factor data are from Refs.\,\cite{Volmer:2000ek, Horn:2006tm, Tadevosyan:2007yd, Blok:2008jy, Huber:2008id}.
}
\end{figure}

This final step is repeated for every member of each DF ensemble.  We thereby arrive at data-driven GPDs specified by Eqs.\,\eqref{hadronscaleDF}, \eqref{eq:GPD}, \eqref{eq:PhiPade} and the parameters in Table~\ref{tab:PhiPade}.
A check on the $F_\pi(\Delta^2)$ part of the procedure is provided by Fig.\,\ref{FigFpi}, which displays the pion electromagnetic form factor obtained via Eq.\,\eqref{eq:sumrule} using these pion GPD ensembles.
Once again, the data-driven results are in accord with modern theory predictions.

\medskip

\begin{figure}[t]
\vspace*{2ex}

\leftline{\hspace*{0.5em}{\large{\textsf{A}}}}
\vspace*{-3ex}
\hspace*{1em}\includegraphics[width=0.435\textwidth]{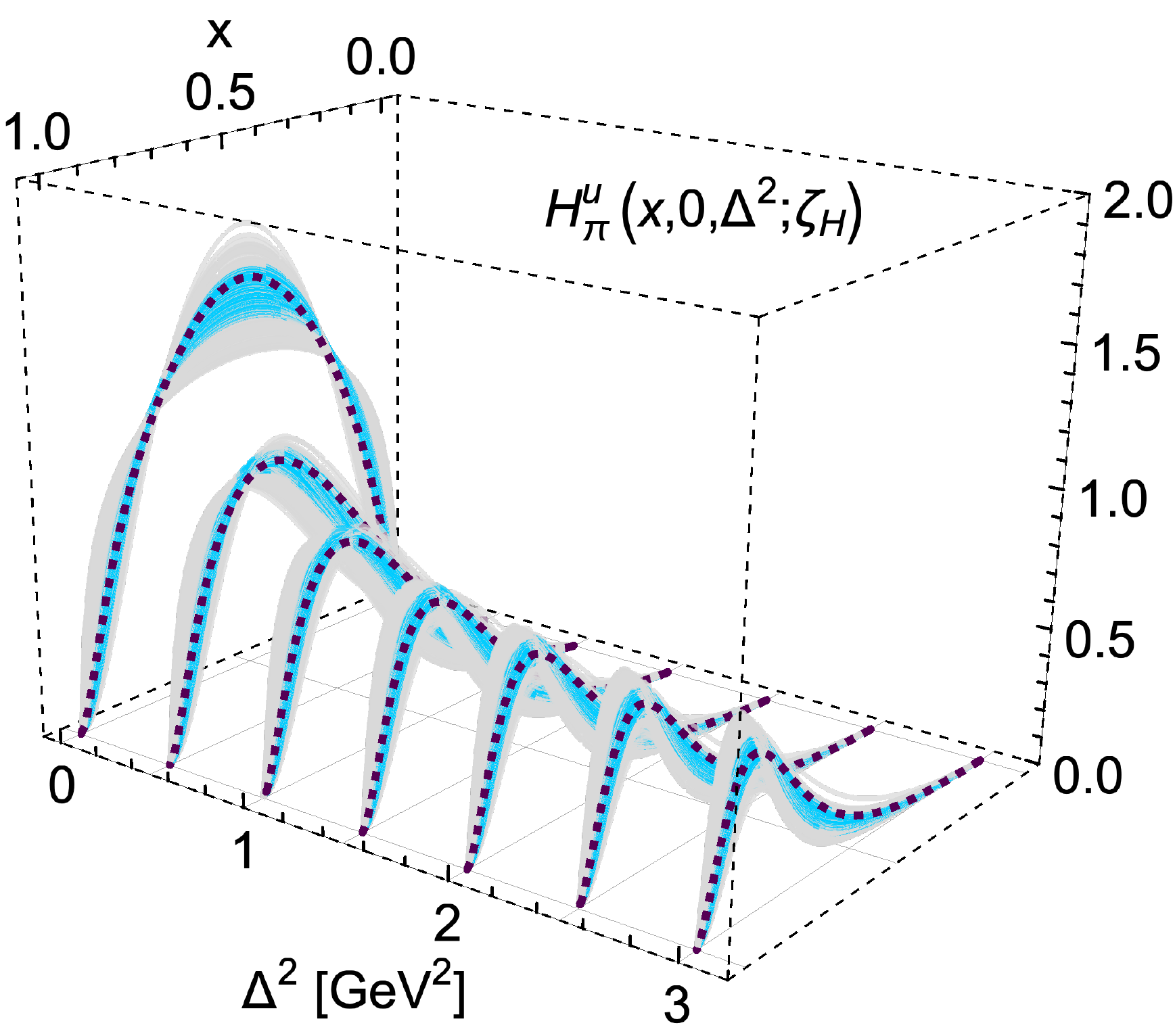}
\vspace*{2ex}

\leftline{\hspace*{0.5em}{\large{\textsf{B}}}}
\vspace*{-3ex}
\hspace*{1em}\includegraphics[width=0.435\textwidth]{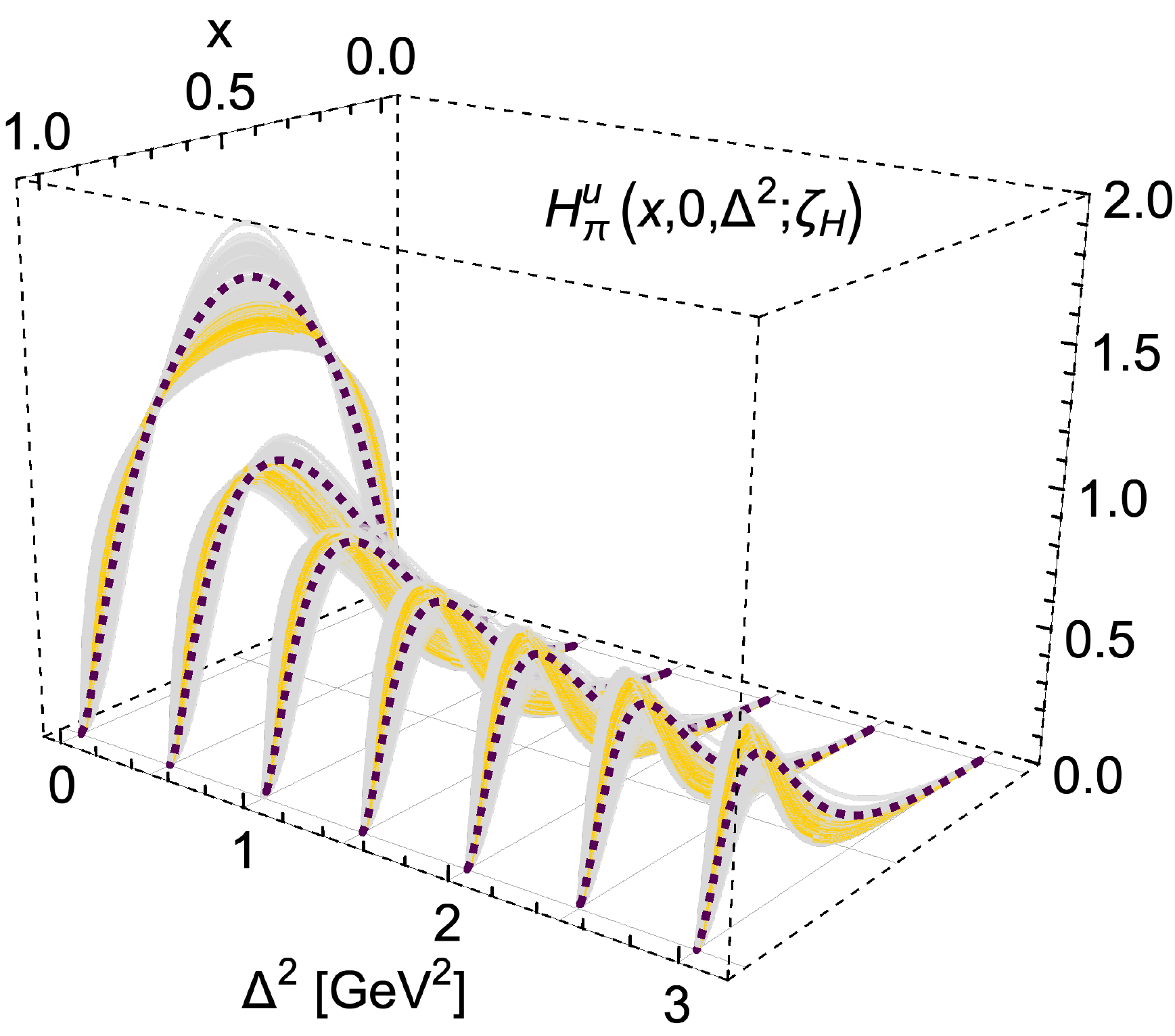}

\caption{
\label{fig:GPDs}
Pion GPDs.
\emph{Panel A}. Working with DFs ${\mathpzc u}^\pi_{\rm A}(x;\zeta_{\cal H})$ \cite{Aicher:2010cb} -- blue band.
\emph{Panel B}. Using DFs ${\mathpzc u}^\pi_{\rm B}(x;\zeta_{\cal H})$ \cite[Sec.\,8]{Cui:2021mom} -- orange band.
Comparison curves, both panels:
CSM prediction in Refs.\,\cite{Zhang:2021mtn, Raya:2021zrz} -- dashed purple curve;
GPD ensemble generated from valence-quark DFs developed in Ref.\,\cite{Cui:2022bxn}, obtained from results computed using lattice Schwinger function methods \cite{Joo:2019bzr, Sufian:2019bol, Alexandrou:2021mmi} -- grey band.
}
\end{figure}

\noindent\emph{4.$\;$Pion Generalised Parton Distribution} ---
Pion GPDs, reconstructed, as described above, from available analyses of relevant Drell-Yan and electron+pion scattering data, are drawn in Fig.\,\ref{fig:GPDs}.  Although the different ensembles are only marginally compatible with each other, owing to differences between the analyses in Refs.\,\cite{Aicher:2010cb, Barry:2021osv}, they both agree with the lQCD based ensembles, within mutual uncertainties, because the lQCD-constrained ensemble possesses a large uncertainty.  The CSM prediction favours the ${\mathpzc u}^\pi_{\rm A}(x;\zeta_{\cal H})$ ensemble.  In all cases, one sees that the support of the valence-quark GPD becomes increasingly concentrated in the neighbourhood $x\simeq 1$ with increasing $\Delta^2$.  Namely, greater probe momentum focuses attention on the domain in which one valence-quark carries a large fraction of the pion's light-front momentum.

\medskip

\noindent\emph{5.$\;$Pion Mass Distribution} --- The first Mellin moment of the $\xi=0$ GPD is the mass distribution form factor:
\begin{equation}
\label{eq:theta2}
\theta_2^\pi(\Delta^2) =  \int_{-1}^1dx\, 2 x\,H_\pi^{\mathpzc u}(x,0,-\Delta^2;\zeta_{\cal H}) \,,
\end{equation}
which is a principal, dynamical coefficient in the expectation value of the QCD energy-momentum tensor in the pion \cite{Polyakov:2018zvc}.  $\theta_2^\pi(\Delta^2) $ is $\zeta$-independent but the manner by which its strength is shared amongst different parton species does evolve with $\zeta$.  Herein we exploit the fact that, at $\zeta=\zeta_{\cal H}$, $\theta_2^\pi(\Delta^2) $ is completely determined by the contribution from dressed valence degrees-of-freedom.

\begin{figure}[t]
\vspace*{6ex}

\leftline{\hspace*{0.5em}{\large{\textsf{A}}}}
\vspace*{-7ex}
\hspace*{1em}\includegraphics[width=0.41\textwidth]{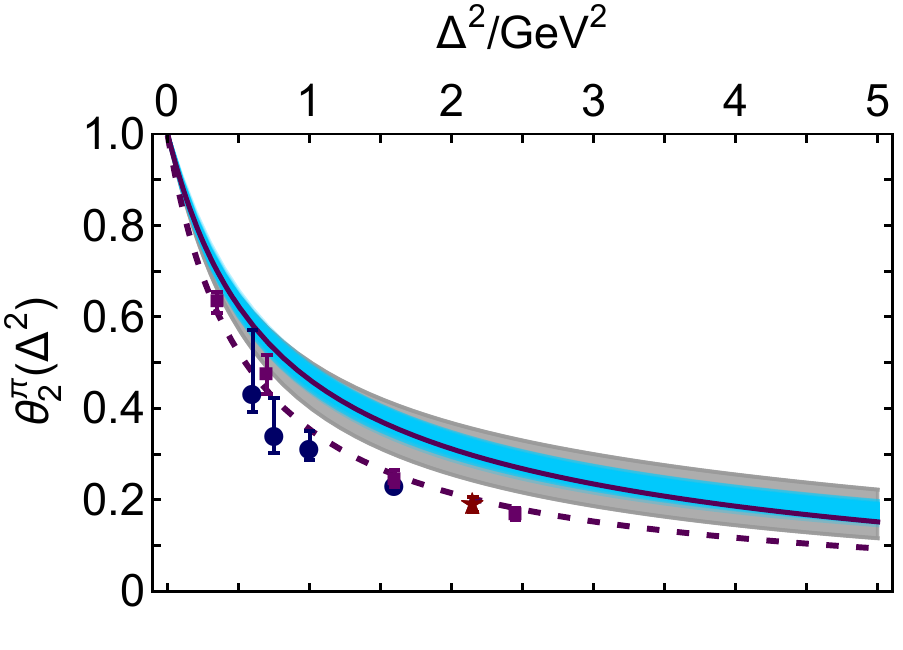}
\vspace*{-4ex}

\leftline{\hspace*{0.5em}{\large{\textsf{B}}}}
\vspace*{-3ex}
\hspace*{1em}\includegraphics[width=0.41\textwidth]{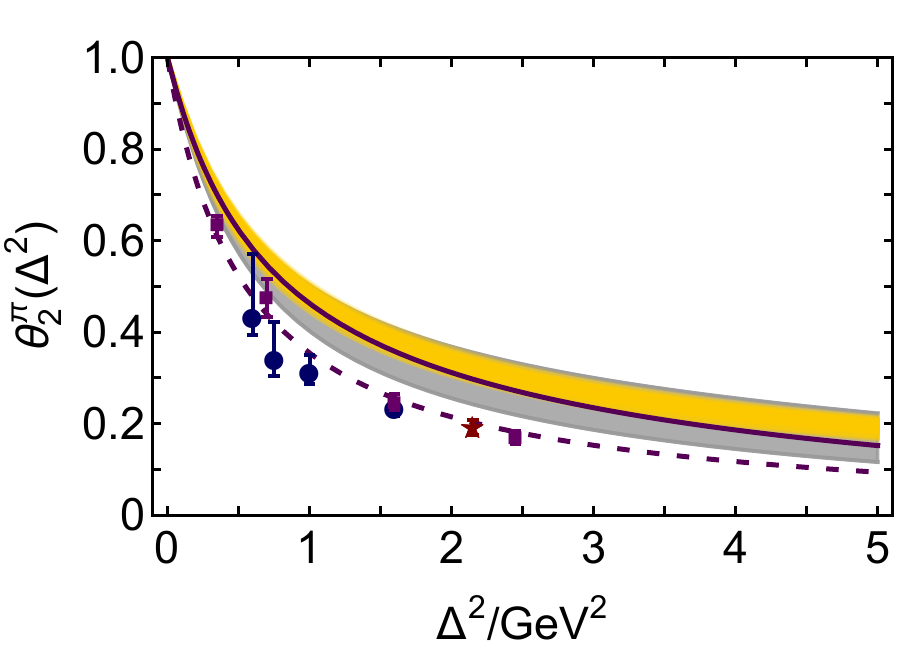}

\caption{\label{fig:FFs}
Pion mass distribution form factor, $\theta_2^\pi(\Delta^2)$.
\emph{Panel A}. Developed from the ${\mathpzc u}^\pi_{\rm A}(x;\zeta_{\cal H})$ ensemble \cite{Aicher:2010cb} -- blue band.
\emph{Panel B}. Developed from the ${\mathpzc u}^\pi_{\rm B}(x;\zeta_{\cal H})$ ensemble \cite[Sec.\,8]{Cui:2021mom} -- orange band.
Comparison curves, both panels:
CSM prediction for $\theta_2^\pi(\Delta^2)$ in Refs.\,\cite{Zhang:2021mtn, Raya:2021zrz} -- solid purple;
GPD ensemble generated from valence-quark DFs developed in Ref.\,\cite{Cui:2022bxn} using lQCD results \cite{Joo:2019bzr, Sufian:2019bol, Alexandrou:2021mmi} -- grey band.
In addition, each panel displays the CSM prediction for $F_\pi(\Delta^2)$ \cite[Sec.\,4B]{Roberts:2021nhw}, \cite{Chen:2018rwz} -- dashed purple curve.
The data are those for $F_\pi(\Delta^2)$ from Refs.\,\cite{Volmer:2000ek, Horn:2006tm, Tadevosyan:2007yd, Blok:2008jy, Huber:2008id}, included so as to highlight the precision required to distinguish the mass and electromagnetic form factors.
}
\end{figure}

The pion mass distributions obtained from the data-driven GPDs are depicted in Fig.\,\ref{fig:FFs}.  Plainly, the mass distribution form factor is harder than the elastic electromagnetic form factor, \emph{i.e}., the distribution of mass in the pion is more compact than the distribution of electric charge.  Since every curve herein is built solely from available data, then this is an empirical fact.  It is also readily understood theoretically, as we now explain.

Beginning with Eqs.\,\eqref{eq:sumrule}, \eqref{eq:theta2}, consider the difference between mass and charge distributions,
\begin{align}
\theta_2^\pi(\Delta^2) & -F_\pi(\Delta^2)  = \nonumber \\
& \int_0^1 d\bar x\, (1-2\bar x)\, {\mathpzc u}^\pi(\bar x;\zeta_{\cal H})\,\Phi^\pi(\Delta^2 \bar x^2;\zeta_H) \,,
\end{align}
where $\bar x = 1-x$ and we have used Eq.\,\eqref{EqSymmetric} to simplify the right-hand side (rhs).
For $\Delta^2=0$, the rhs is zero as a consequence of charge and baryon-number conservation.
Suppose now that $\Phi^\pi(z;\zeta_H)$ is a non-negative, monotonically decreasing function of its argument, as required to produce a realistic pion electromagnetic form factor, then it is straightforward to establish that
\begin{equation}
\forall \Delta^2 > 0 \,,\quad \theta_2^\pi(\Delta^2) -F_\pi(\Delta^2) > 0\,.
\end{equation}
Consequently, the pion's mass distribution is more compact than its charge distribution in any realistic description of pion properties; in particular, the mass radius is smaller than the charge radius:
$(r_\pi^{\theta_2})^2 <r_\pi^2$,
as demonstrated previously in Ref.\,\cite[Eq.\,(41)]{Raya:2021zrz}.

Using Eq.\,\eqref{eq:theta2} and the ensemble characterisation parameters in Table~\ref{tab:PhiPade}, one finds (in fm):
\begin{equation}
\label{MassRadii}
\begin{array}{l|ccc}
 & {\rm A} & {\rm B} & {\rm lQCD} \\\hline
r_\pi^{\theta_2} & 0.518(16) & 0.498(14) & 0.512(21)
\end{array}\,,
\end{equation}
to be compared with $r_\pi/{\rm fm} = 0.64(2)$.  These results yield the data-driven prediction $r_\pi^{\theta_2} /r_\pi = 0.79(3)$, which translates into a volume ratio of $0.49(6)$.

Considering Ref.\,\cite[Eq.\,(41)]{Raya:2021zrz}, it is plain that the radii and associated uncertainties in Eq.\,\eqref{MassRadii} are determined by knowledge of $r_\pi$ and the DF ensembles.
This simplicity is a feature of the factorised \emph{Ansatz} for the LFWF.
As evident from Ref.\,\cite[Fig.\,6A]{Raya:2021zrz}, on the domain $\Delta^2 r_\pi^2 \simeq 0$, relevant to radius determination, there is practically no difference between pion GPDs obtained using factorised and nonfactorised LFWFs.  Hence, in the context of Eq.\,\eqref{MassRadii}, the factorised representation is a reliable approximation.

Exploiting these features further and adapting the arguments that lead to Ref.\,\cite[Eq.\,(7)]{Cui:2022bxn}, one arrives at the following constraints on the radii ratio for pion-like bound-states:
\begin{equation}
\label{bounds}
\tfrac{1}{\surd 2} \leq r_\pi^{\theta_2} /r_\pi \leq 1\,,
\end{equation}
where the lower bound is saturated by a point-particle DF,
${\mathpzc u}(x;\zeta_{\cal H})=\theta(x)\theta(1-x)$,
and the upper by the DF of a bound-state formed from infinitely massive valence degrees-of-freedom, ${\mathpzc u}(x;\zeta_{\cal H})=\delta(x-1/2)$.


In the context of Eqs.\,\eqref{MassRadii}, \eqref{bounds}, we note that an analysis of $\gamma^\ast \gamma \to \pi^0 \pi^0$ data \cite{Belle:2015oin} using a generalised distribution amplitude formalism, with subsequent analytic continuation to spacelike momenta to enable extraction of a mass radius, yields \cite{Kumano:2017lhr}:
$r_\pi^{\theta_2}/r_\pi = 0.49 - 0.59$.
Although the radii size ordering is qualitatively similar to our determination, the value of the ratio is smaller and inconsistent with the bounds in Eq.\,\eqref{bounds}.
Furthermore, regarding theory, CSMs produce the results drawn in Figs.\,\ref{FigFpi}, \ref{fig:FFs}, which yield
$r_\pi^{\theta_2} /r_\pi = 0.81(3)$, \emph{viz}.\ a value consistent with the data-driven result.
On the other hand, analyses based on momentum-independent quark+antiquark interactions produce smaller values: $r_\pi^{\theta_2} /r_\pi = 0.71$ \cite{Shastry:2022obb} and $r_\pi^{\theta_2} /r_\pi = 0.53$ \cite{Xing:2022mvk}.  Here, for reasons explained elsewhere \cite{Chen:2012txa}, the mutual inconsistency is attributable to the different treatments of the momentum-independent interaction used in those studies.
Moreover, the Ref.\,\cite{Shastry:2022obb} result is consistent with Eq.\,\eqref{bounds} because it omits resonance pole contributions to both radii and that in Ref.\,\cite{Xing:2022mvk} fails the test because it only includes such contributions to $r_\pi$.

It is worth recording here that the pion LFWF, hence GPD, is independent of the probe: it is the same whether a photon or graviton is the probing object.  However, the probe itself is sensitive to different features of the target constituents.  A target dressed-quark carries the same charge, irrespective of its momentum.  So, the pion wave function alone controls the distribution of charge.  On the other hand, the gravitational interaction of a target dressed-quark depends on its momentum -- see Eq.\,\eqref{eq:theta2}; after all, the current relates to the energy momentum tensor.  The pion effective mass distribution therefore depends on interference between quark momentum growth and LFWF momentum suppression.  This pushes support to a larger momentum domain in the pion, which corresponds to a smaller distance domain.

\medskip

\noindent\emph{6.$\;$GPD in Impact Parameter Space} --- An impact parameter space (IPS) GPD is defined via a Hankel transform:
\begin{equation}
{\mathpzc u}^{\pi}(x,b_\perp^2;\zeta_{\mathpzc H}) = \int_0^\infty \frac{d\Delta}{2\pi} \Delta J_0(|b_\perp| \Delta) \, H_{\pi}^u(x,0,-\Delta^2;\zeta_{\mathpzc H})\,,
\label{eq:IPDHgen}
\end{equation}
where $J_0$ is a cylindrical Bessel function.  This is a density, which reveals the probability of finding a valence-quark within the light-front at a transverse distance $|b_\perp|$ from the meson's centre of transverse momentum.  Using Eq.\,\eqref{eq:GPD}, Eq.\,\eqref{eq:IPDHgen} simplifies:\footnote{Motivated by the presence of scaling violations in $F_\pi(\Delta^2)$ \cite{Lepage:1979za, Efremov:1979qk, Lepage:1980fj} and informed by Refs.\,\cite[Eq.\,(38)]{Raya:2021zrz} \cite[Eqs.\,(14)]{Qin:2017lcd}, we ensure the Hankel transforms are finite by matching each member of a form factor ensemble to ($y=\Delta^2$) $[(1+a_1 y)/(1+a_2 y + (a_3 y)^2)][(1+ a_0 y)/(1+a_0 y \ln[1+a_0 y])]$, with $a_{0,1,2,3}$ being fitting parameters.}
\begin{subequations}
\label{eq:IPSfac}
\begin{align}
{\mathpzc u}^\pi(x,b_\perp^2;\zeta_H) &= \frac{{\mathpzc u}^\pi(x;\zeta_H)}{(1-x)^2} \Psi^\pi\left(\frac{|b_\perp|}{1-x};\zeta_H \right) \,, \label{eq:IPSfac1} \\
\Psi^\pi(u;\zeta_H) &=  \int_0^\infty \frac{ds}{2\pi} s \, J_0(u s)
\, \Phi^\pi(s^2;\zeta_H) \,.
 \label{eq:IPSfac2}
\end{align}
\end{subequations}
Considering the character of the entries in Eqs.\,\eqref{eq:IPSfac}, it becomes clear that the global maximum of the IPS GPD is given by the value of ${\mathpzc u}^{\pi} (x=1, b_\perp^2=0;\zeta_{\mathpzc H})$.
\begin{figure}[t]
\vspace*{2ex}

\leftline{\hspace*{0.5em}{\large{\textsf{A}}}}
\vspace*{-3ex}
\hspace*{1em}\includegraphics[width=0.435\textwidth]{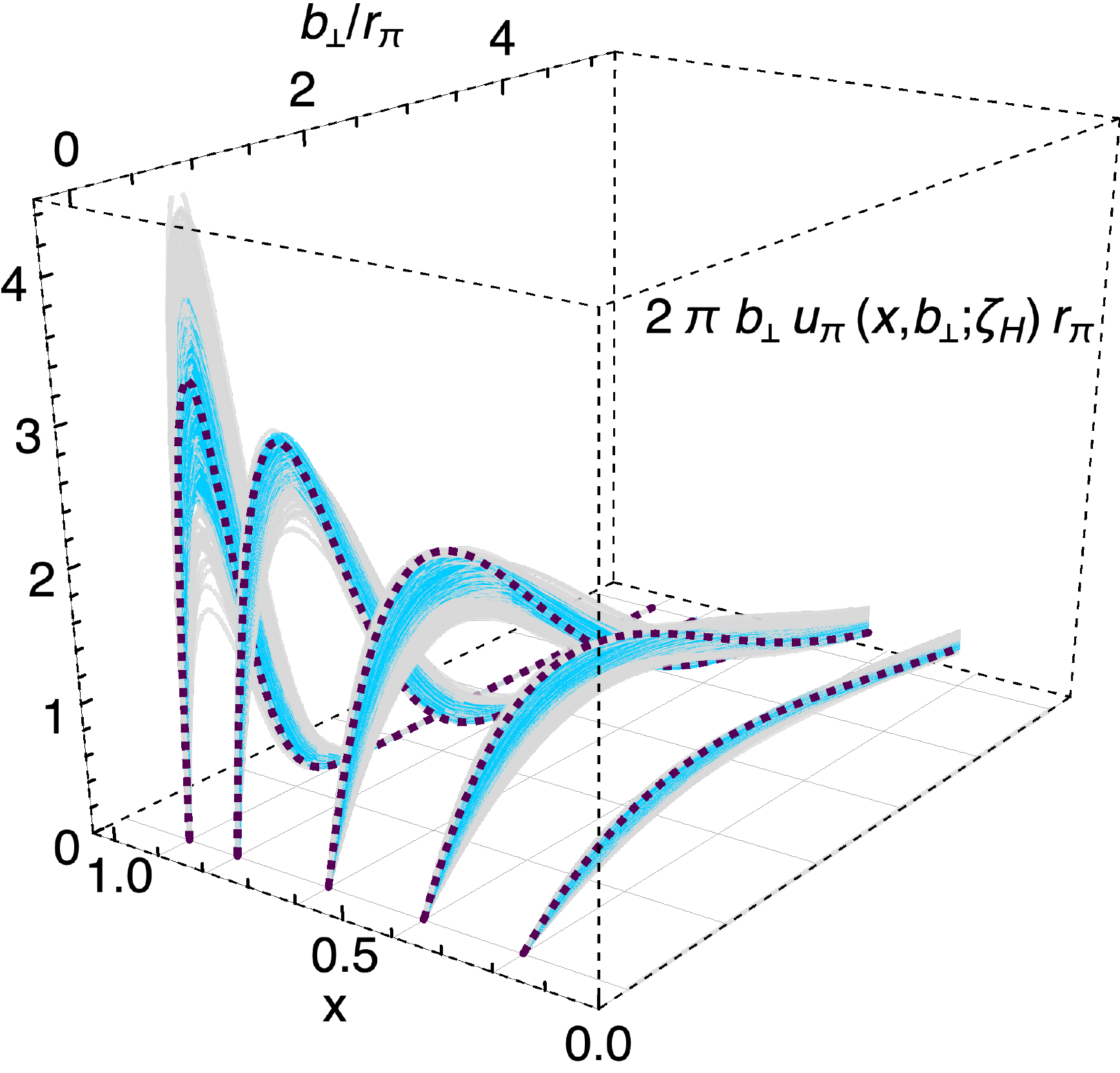}
\vspace*{2ex}

\leftline{\hspace*{0.5em}{\large{\textsf{B}}}}
\vspace*{-3ex}
\hspace*{1em}\includegraphics[width=0.435\textwidth]{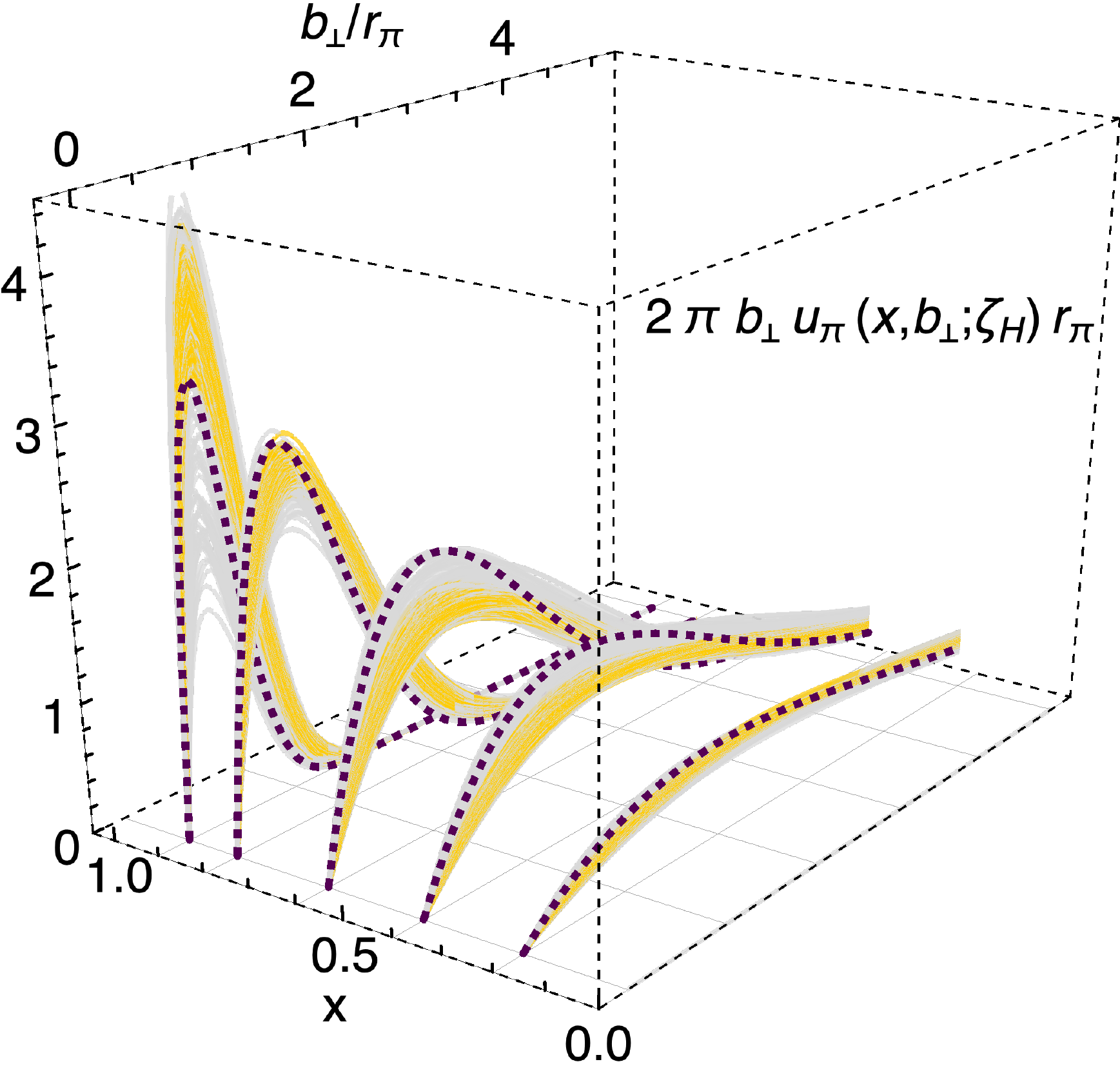}

\caption{
Pion GPDs in impact parameter space, displayed as functions of $|b_\perp|$ at $x=0.9, 0.8, 0.6, 0.4, 0.2, 0$.
\emph{Panel A}. Working with DFs ${\mathpzc u}^\pi_{\rm A}(x;\zeta_{\cal H})$ \cite{Aicher:2010cb} -- blue band.
\emph{Panel B}. Using DFs ${\mathpzc u}^\pi_{\rm B}(x;\zeta_{\cal H})$ \cite[Sec.\,8]{Cui:2021mom} -- orange band.
Comparison curves, both panels:
CSM prediction in Refs.\,\cite{Zhang:2021mtn, Raya:2021zrz} -- dashed purple curve;
GPD ensemble generated from valence-quark DFs developed in Ref.\,\cite{Cui:2022bxn}, obtained from results computed using lattice Schwinger function methods \cite{Joo:2019bzr, Sufian:2019bol, Alexandrou:2021mmi} -- grey band.
\label{fig:ips}}
\end{figure}

In depicting the IPS GPD, it is usual to draw $2\pi |b_\perp| {\mathpzc u}^{\pi}( x, b_\perp^2;\zeta_{\mathpzc H})$. The peak in this function is shifted to $|b_\perp|>0$ by an amount that reflects aspects of bound-state structure.
These features are evident in Fig.\,\ref{fig:ips}, which displays the IPS GPDs obtained from both A and B ensembles.
%
The $x>0$ locations of the global maxima and their intensities, given as a triplet $\{x,b_\perp/r_\pi,\mathpzc{i}_\pi\}$, are:
%
\begin{equation}
\begin{array}{l|ccc}
   & x & b_\perp/r_\pi & \mathpzc{i}_\pi \\\hline
{\rm CSM}\, \mbox{\cite{Raya:2021zrz}} & 0.88\phantom{(6)} & 0.13\phantom{(6)} & 3.29\phantom{(1.67)}\\
{\rm A} & 0.89(2) &  0.10(2) &  3.21(30) \phantom{.1}\\
{\rm B} & 0.95(1) &  0.05(1) &  4.58(50) \phantom{.1} \\
{\rm lQCD} & 0.91(6) &  0.08(5) & 4.04(1.67)
\end{array}\,.
\end{equation}
The valence IPS GPDs become increasingly broad as $x\to 0$ because, with diminishing $x$, the valence degree-of-freedom plays a progressively smaller role in defining the pion's centre of transverse momentum.

\begin{figure}[t]
\vspace*{6ex}

\leftline{\hspace*{0.5em}{\large{\textsf{A}}}}
\vspace*{-7ex}
\hspace*{1em}\includegraphics[width=0.41\textwidth]{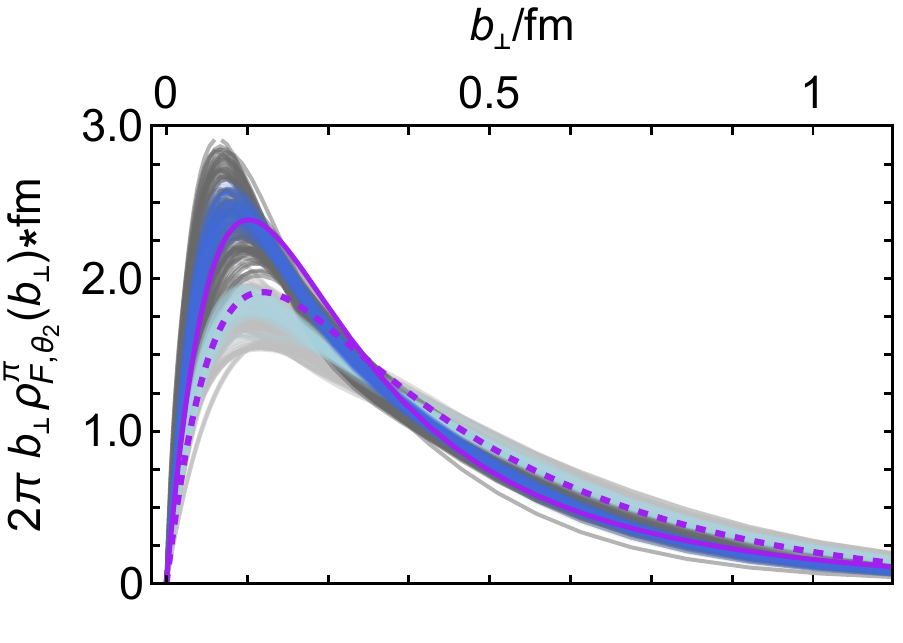}
\vspace*{-4ex}

\leftline{\hspace*{0.5em}{\large{\textsf{B}}}}
\vspace*{-3ex}
\hspace*{1em}\includegraphics[width=0.41\textwidth]{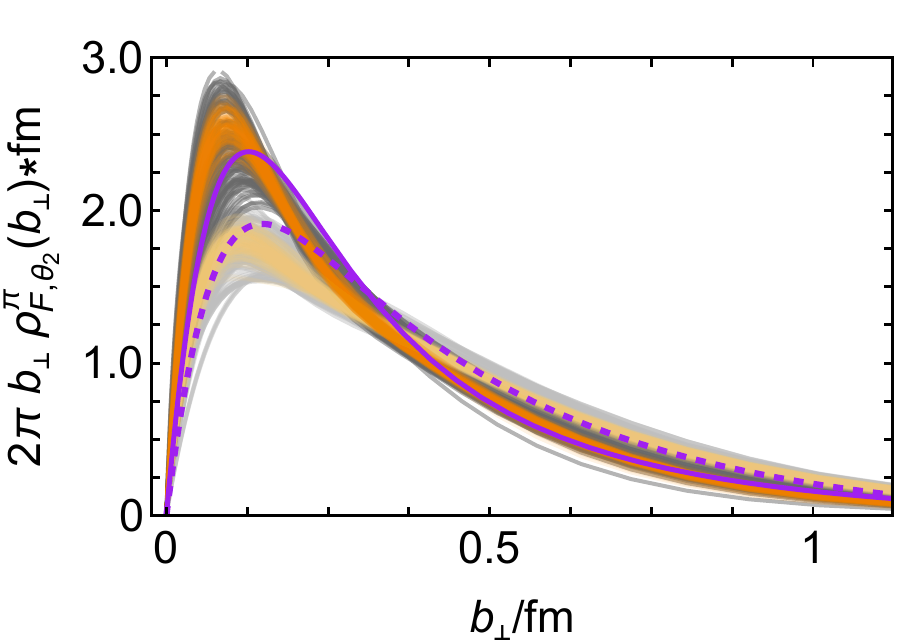}

\caption{\label{fig:dists}
Light-front transverse density distributions, Eqs.\,\eqref{DensityLFT}, built from: ${\mathpzc u}^\pi_{\rm A}(x;\zeta_{\cal H})$ ensemble \cite{Aicher:2010cb} -- \emph{Panel A} (charge $=$ light blue, mass $=$ dark blue) ; and ${\mathpzc u}^\pi_{\rm B}(x;\zeta_{\cal H})$ ensemble \cite[Sec.\,8]{Cui:2021mom} -- \emph{Panel B} (charge $=$ light orange, mass $=$ dark orange).
Comparison curves, both panels:
dashed purple curve -- charge density calculated using CSM prediction for $F_\pi(\Delta^2)$ \cite[Sec.\,4B]{Roberts:2021nhw}, \cite{Chen:2018rwz};
solid purple curve -- mass density obtained using CSM prediction for $\theta_2^\pi(\Delta^2)$ in Refs.\,\cite{Zhang:2021mtn, Raya:2021zrz};
silver band -- lattice-based charge density ensemble;
and
grey band -- lattice-based mass density ensemble.
}
\end{figure}

Density distributions in the light-front transverse plane are obtained by integrating a given IPS GPD and its $x$-weighted partner over the light-front momentum fraction:
%
{\allowdisplaybreaks
\begin{align}
\rho_{\{F,\theta_2\} }^\pi & (|b_\perp|)  = \int_{-1}^1 dx\,
\{1,2x \} {\mathpzc u}^{\pi}(x,b_\perp^2;\zeta_{\mathpzc H}) \nonumber \\
& = \int_0^\infty \frac{d\Delta}{2\pi} \Delta J_0(|b_\perp|\Delta) \{ F_\pi(\Delta^2),\theta_2(\Delta^2) \} \,.
\label{DensityLFT}
\end{align}
Here we have used isospin symmetry to identify the charge density associated with a given valence-quark flavour with the net pion density.  Again, these quantities are $\zeta$-independent.}

The light-front transverse densities in Eqs.\,\eqref{DensityLFT} are drawn in Fig.\,\ref{fig:dists}.  Evidently, consistent with the analysis presented above, the pion's light-front transverse mass distribution is more compact than the analogous charge distribution.  Moreover, the data-built results are consistent with modern theory predictions.

\medskip

\noindent\emph{7.$\;$Summary and Perspective} ---
Supposing that there exists an effective charge which defines an evolution scheme for parton distribution functions (DFs) that is all-orders exact \cite{Cui:2019dwv, Cui:2021mom, Cui:2022bxn} and working solely with existing pion+nucleus Drell-Yan and electron+pion scattering data  \cite{Conway:1989fs, Amendolia:1984nz, Amendolia:1986wj, Volmer:2000ek, Horn:2006tm, Tadevosyan:2007yd, Blok:2008jy, Huber:2008id}, we used a $\chi^2$-based selection procedure to develop ensembles of model-independent representations of the three-dimensional pointwise behaviour of the pion generalised parton distribution (GPD).  These ensembles yield a data-driven prediction for the pion mass distribution form factor, $\theta_2$.
In comparison with the pion elastic electromagnetic form factor, obtained analogously, $\theta_2$ is more compact; and based on extant data, the ratio of the radii derived from these form factors is $r_\pi^{\theta_2}/r_\pi = 0.79(3)$.

Our results are reported at the hadron scale, $\zeta_{\cal H}$, whereat, by definition, all hadron properties are carried by dressed valence quasiparticle degrees-of-freedom.  Nevertheless, using the all-orders evolution scheme, one can readily determine the manner in which these distributions are shared amongst the various parton species at any scale $\zeta>\zeta_{\cal H}$ -- see, \emph{e.g}., Ref.\,\cite[Sec.\,VIII]{Raya:2021zrz}.

Improvement of our data-driven predictions must await new data relevant to pion DFs and improvements in associated analysis methods; and pion form factor data that extends to larger momentum transfers than currently available.
Meanwhile, our results for the pion GPD, related form factors and distributions should serve as valuable constraints on modern pion structure theory.

No similar analysis for the kaon will be possible before analogous empirical information becomes available.  In this case, today, the kaon charge radius cannot be considered known \cite{Cui:2021aee} because kaon elastic form factor data are sketchy \cite{Dally:1980dj, Amendolia:1986ui} and only eight data relevant to kaon valence-quark DFs are available \cite{Badier:1980jq}.

Given the importance of contrasting pion and proton mass distributions in the search for an understanding of emergent hadron mass, completing a kindred analysis that leads to data-driven ensembles of proton GPDs should be given high priority.  Precise elastic form factor data are available \cite{Punjabi:2015bba, Gao:2021sml, Cui:2022fyr}; however, despite a wealth of data relevant to proton DFs, improved analyses are required, including, \emph{e.g}., effects of next-to-leading-logarithm resummation.  Meanwhile, one might begin with existing theory predictions that provide a unified description of pion and proton DFs \cite{Lu:2022cjx}.

%
\medskip
\noindent\emph{Acknowledgments}.
%
We are grateful for constructive comments from G.\,M.~Huber.
Work supported by:
National Natural Science Foundation of China (grant nos.\,12135007, 12233002);
%
Natural Science Foundation of Jiangsu Province (grant no.\ BK20220122);
Spanish Ministry of Science and Innovation (MICINN grant no.\ PID2019-107844GB-C22);
and
Junta de Andaluc{\'{\i}}a (grant no.\ P18-FR-5057).


\begin{thebibliography}{90}
\providecommand{\natexlab}[1]{#1}
\providecommand{\url}[1]{\texttt{#1}}
\providecommand{\urlprefix}{URL }
\expandafter\ifx\csname urlstyle\endcsname\relax
  \providecommand{\doi}[1]{doi:\discretionary{}{}{}#1}\else
  \providecommand{\doi}[1]{doi:\discretionary{}{}{}\begingroup
  \urlstyle{rm}\url{#1}\endgroup}\fi
\providecommand{\bibinfo}[2]{#2}

\bibitem[{Adams et~al.(2018)}]{Adams:2018pwt}
\bibinfo{author}{B.~Adams}, et~al., \bibinfo{title}{{Letter of Intent: A New
  QCD facility at the M2 beam line of the CERN SPS (COMPASS++/AMBER) --
  arXiv:1808.00848 [hep-ex]$\!$}} .

\bibitem[{Quintans(2022)}]{Quintans:2022utc}
\bibinfo{author}{C.~Quintans}, \bibinfo{title}{{The New AMBER Experiment at the
  CERN SPS}}, \bibinfo{journal}{Few Body Syst.}
  \bibinfo{volume}{63}~(\bibinfo{number}{4}) (\bibinfo{year}{2022})
  \bibinfo{pages}{72}.

\bibitem[{Aguilar et~al.(2019)}]{Aguilar:2019teb}
\bibinfo{author}{A.~C. Aguilar}, et~al., \bibinfo{title}{{Pion and Kaon
  Structure at the Electron-Ion Collider}}, \bibinfo{journal}{Eur. Phys. J. A}
  \bibinfo{volume}{55} (\bibinfo{year}{2019}) \bibinfo{pages}{190}.

\bibitem[{Brodsky et~al.(2020)}]{Brodsky:2020vco}
\bibinfo{author}{S.~J. Brodsky}, et~al., \bibinfo{title}{{Strong QCD from
  Hadron Structure Experiments}}, \bibinfo{journal}{Int. J. Mod. Phys. E}
  \bibinfo{volume}{29}~(\bibinfo{number}{08}) (\bibinfo{year}{2020})
  \bibinfo{pages}{2030006}.

\bibitem[{Chen et~al.(2020)Chen, Guo, Roberts, and Wang}]{Chen:2020ijn}
\bibinfo{author}{X.~Chen}, \bibinfo{author}{F.-K. Guo}, \bibinfo{author}{C.~D.
  Roberts}, \bibinfo{author}{R.~Wang}, \bibinfo{title}{{Selected Science
  Opportunities for the EicC}}, \bibinfo{journal}{Few Body Syst.}
  \bibinfo{volume}{61} (\bibinfo{year}{2020}) \bibinfo{pages}{43}.

\bibitem[{Anderle et~al.(2021)}]{Anderle:2021wcy}
\bibinfo{author}{D.~P. Anderle}, et~al., \bibinfo{title}{{Electron-ion collider
  in China}}, \bibinfo{journal}{Front. Phys. (Beijing)}
  \bibinfo{volume}{16}~(\bibinfo{number}{6}) (\bibinfo{year}{2021})
  \bibinfo{pages}{64701}.

\bibitem[{Arrington et~al.(2021)}]{Arrington:2021biu}
\bibinfo{author}{J.~Arrington}, et~al., \bibinfo{title}{{Revealing the
  structure of light pseudoscalar mesons at the electron\textendash{}ion
  collider}}, \bibinfo{journal}{J. Phys. G} \bibinfo{volume}{48}
  (\bibinfo{year}{2021}) \bibinfo{pages}{075106}.

\bibitem[{Abdul~Khalek et~al.(2022)}]{AbdulKhalek:2021gbh}
\bibinfo{author}{R.~Abdul~Khalek}, et~al., \bibinfo{title}{{Science
  Requirements and Detector Concepts for the Electron-Ion Collider: EIC Yellow
  Report}}, \bibinfo{journal}{Nucl. Phys. A} \bibinfo{volume}{1026}
  (\bibinfo{year}{2022}) \bibinfo{pages}{122447}.

\bibitem[{Wang and Chen(2022)}]{Wang:2022xad}
\bibinfo{author}{R.~Wang}, \bibinfo{author}{X.~Chen}, \bibinfo{title}{{The
  Current Status of Electron Ion Collider in China}}, \bibinfo{journal}{Few
  Body Syst.} \bibinfo{volume}{63}~(\bibinfo{number}{2}) (\bibinfo{year}{2022})
  \bibinfo{pages}{48}.

\bibitem[{Carman et~al.(2023)Carman, Gothe, Mokeev, and
  Roberts}]{Carman:2023zke}
\bibinfo{author}{D.~S. Carman}, \bibinfo{author}{R.~W. Gothe},
  \bibinfo{author}{V.~I. Mokeev}, \bibinfo{author}{C.~D. Roberts},
  \bibinfo{title}{{Nucleon Resonance Electroexcitation Amplitudes and Emergent
  Hadron Mass\,--\,arXiv:2301.07777 [hep-ph]}}, \bibinfo{journal}{Particles}
  \bibinfo{volume}{6}~(\bibinfo{number}{3}) (\bibinfo{year}{2023})
  \bibinfo{pages}{\emph{in press}}.

\bibitem[{Machleidt and Entem(2011)}]{Machleidt:2011zz}
\bibinfo{author}{R.~Machleidt}, \bibinfo{author}{D.~R. Entem},
  \bibinfo{title}{{Chiral effective field theory and nuclear forces}},
  \bibinfo{journal}{Phys. Rept.} \bibinfo{volume}{503} (\bibinfo{year}{2011})
  \bibinfo{pages}{1--75}.

\bibitem[{Aad et~al.(2012)}]{Aad:2012tfa}
\bibinfo{author}{G.~Aad}, et~al., \bibinfo{title}{{Observation of a new
  particle in the search for the Standard Model Higgs boson with the ATLAS
  detector at the LHC}}, \bibinfo{journal}{Phys. Lett. B} \bibinfo{volume}{716}
  (\bibinfo{year}{2012}) \bibinfo{pages}{1--29}.

\bibitem[{Chatrchyan et~al.(2012)}]{Chatrchyan:2012xdj}
\bibinfo{author}{S.~Chatrchyan}, et~al., \bibinfo{title}{{Observation of a New
  Boson at a Mass of 125 GeV with the CMS Experiment at the LHC}},
  \bibinfo{journal}{Phys. Lett. B} \bibinfo{volume}{716} (\bibinfo{year}{2012})
  \bibinfo{pages}{30--61}.

\bibitem[{Englert(2014)}]{Englert:2014zpa}
\bibinfo{author}{F.~Englert}, \bibinfo{title}{{Nobel Lecture: The BEH mechanism
  and its scalar boson}}, \bibinfo{journal}{Rev. Mod. Phys.}
  \bibinfo{volume}{86} (\bibinfo{year}{2014}) \bibinfo{pages}{843}.

\bibitem[{Higgs(2014)}]{Higgs:2014aqa}
\bibinfo{author}{P.~W. Higgs}, \bibinfo{title}{{Nobel Lecture: Evading the
  Goldstone theorem}}, \bibinfo{journal}{Rev. Mod. Phys.} \bibinfo{volume}{86}
  (\bibinfo{year}{2014}) \bibinfo{pages}{851}.

\bibitem[{Roberts(2017)}]{Roberts:2016vyn}
\bibinfo{author}{C.~D. Roberts}, \bibinfo{title}{{Perspective on the origin of
  hadron masses}}, \bibinfo{journal}{Few Body Syst.} \bibinfo{volume}{58}
  (\bibinfo{year}{2017}) \bibinfo{pages}{5}.

\bibitem[{Roberts and Schmidt(2020)}]{Roberts:2020udq}
\bibinfo{author}{C.~D. Roberts}, \bibinfo{author}{S.~M. Schmidt},
  \bibinfo{title}{{Reflections upon the Emergence of Hadronic Mass}},
  \bibinfo{journal}{Eur. Phys. J. ST}
  \bibinfo{volume}{229}~(\bibinfo{number}{22-23}) (\bibinfo{year}{2020})
  \bibinfo{pages}{3319--3340}.

\bibitem[{Roberts(2020)}]{Roberts:2020hiw}
\bibinfo{author}{C.~D. Roberts}, \bibinfo{title}{{Empirical Consequences of
  Emergent Mass}}, \bibinfo{journal}{Symmetry} \bibinfo{volume}{12}
  (\bibinfo{year}{2020}) \bibinfo{pages}{1468}.

\bibitem[{Krein and Peixoto(2020)}]{Krein:2020yor}
\bibinfo{author}{G.~Krein}, \bibinfo{author}{T.~C. Peixoto},
  \bibinfo{title}{{Femtoscopy of the Origin of the Nucleon Mass}},
  \bibinfo{journal}{Few Body Syst.} \bibinfo{volume}{61}~(\bibinfo{number}{4})
  (\bibinfo{year}{2020}) \bibinfo{pages}{49}.

\bibitem[{Roberts(2021)}]{Roberts:2021xnz}
\bibinfo{author}{C.~D. Roberts}, \bibinfo{title}{{On Mass and Matter}},
  \bibinfo{journal}{AAPPS Bulletin} \bibinfo{volume}{31} (\bibinfo{year}{2021})
  \bibinfo{pages}{6}.

\bibitem[{Roberts et~al.(2021)Roberts, Richards, Horn, and
  Chang}]{Roberts:2021nhw}
\bibinfo{author}{C.~D. Roberts}, \bibinfo{author}{D.~G. Richards},
  \bibinfo{author}{T.~Horn}, \bibinfo{author}{L.~Chang},
  \bibinfo{title}{{Insights into the emergence of mass from studies of pion and
  kaon structure}}, \bibinfo{journal}{Prog. Part. Nucl. Phys.}
  \bibinfo{volume}{120} (\bibinfo{year}{2021}) \bibinfo{pages}{103883}.

\bibitem[{Binosi(2022)}]{Binosi:2022djx}
\bibinfo{author}{D.~Binosi}, \bibinfo{title}{{Emergent Hadron Mass in Strong
  Dynamics}}, \bibinfo{journal}{Few Body Syst.}
  \bibinfo{volume}{63}~(\bibinfo{number}{2}) (\bibinfo{year}{2022})
  \bibinfo{pages}{42}.

\bibitem[{Papavassiliou(2022)}]{Papavassiliou:2022wrb}
\bibinfo{author}{J.~Papavassiliou}, \bibinfo{title}{{Emergence of mass in the
  gauge sector of QCD}}, \bibinfo{journal}{Chin. Phys. C}
  \bibinfo{volume}{46}~(\bibinfo{number}{11}) (\bibinfo{year}{2022})
  \bibinfo{pages}{112001}.

\bibitem[{Ding et~al.(2023)Ding, Roberts, and Schmidt}]{Ding:2022ows}
\bibinfo{author}{M.~Ding}, \bibinfo{author}{C.~D. Roberts},
  \bibinfo{author}{S.~M. Schmidt}, \bibinfo{title}{{Emergence of Hadron Mass
  and Structure}}, \bibinfo{journal}{Particles}
  \bibinfo{volume}{6}~(\bibinfo{number}{1}) (\bibinfo{year}{2023})
  \bibinfo{pages}{57--120}.

\bibitem[{Roberts(2022)}]{Roberts:2022rxm}
\bibinfo{author}{C.~D. Roberts}, \bibinfo{title}{{Origin of the Proton Mass --
  arXiv:2211.09905 [hep-ph]}}, \bibinfo{year}{2022}.

\bibitem[{Ferreira and Papavassiliou(2023)}]{Ferreira:2023fva}
\bibinfo{author}{M.~N. Ferreira}, \bibinfo{author}{J.~Papavassiliou},
  \bibinfo{title}{{Gauge Sector Dynamics in QCD}}, \bibinfo{journal}{Particles}
  \bibinfo{volume}{6}~(\bibinfo{number}{1}) (\bibinfo{year}{2023})
  \bibinfo{pages}{312--363}.

\bibitem[{Bhagwat et~al.(2007)Bhagwat, Chang, Liu, Roberts, and
  Tandy}]{Bhagwat:2007ha}
\bibinfo{author}{M.~S. Bhagwat}, \bibinfo{author}{L.~Chang},
  \bibinfo{author}{Y.-X. Liu}, \bibinfo{author}{C.~D. Roberts},
  \bibinfo{author}{P.~C. Tandy}, \bibinfo{title}{{Flavour symmetry breaking and
  meson masses}}, \bibinfo{journal}{Phys. Rev. C} \bibinfo{volume}{76}
  (\bibinfo{year}{2007}) \bibinfo{pages}{045203}.

\bibitem[{Brodsky et~al.(2010)Brodsky, Roberts, Shrock, and
  Tandy}]{Brodsky:2010xf}
\bibinfo{author}{S.~J. Brodsky}, \bibinfo{author}{C.~D. Roberts},
  \bibinfo{author}{R.~Shrock}, \bibinfo{author}{P.~C. Tandy},
  \bibinfo{title}{{New perspectives on the quark condensate}},
  \bibinfo{journal}{Phys. Rev. C} \bibinfo{volume}{82} (\bibinfo{year}{2010})
  \bibinfo{pages}{022201(R)}.

\bibitem[{Qin et~al.(2014)Qin, Roberts, and Schmidt}]{Qin:2014vya}
\bibinfo{author}{S.-X. Qin}, \bibinfo{author}{C.~D. Roberts},
  \bibinfo{author}{S.~M. Schmidt}, \bibinfo{title}{{Ward-Green-Takahashi
  identities and the axial-vector vertex}}, \bibinfo{journal}{Phys. Lett. B}
  \bibinfo{volume}{733} (\bibinfo{year}{2014}) \bibinfo{pages}{202--208}.

\bibitem[{Corden et~al.(1980)}]{Corden:1980xf}
\bibinfo{author}{M.~Corden}, et~al., \bibinfo{title}{{Production of Muon Pairs
  in the Continuum Region by 39.5-{GeV}/c $\pi^\pm$, $K^\pm$, $p$ and $\bar{p}$
  Beams Incident on a Tungsten Target}}, \bibinfo{journal}{Phys. Lett. B}
  \bibinfo{volume}{96} (\bibinfo{year}{1980}) \bibinfo{pages}{417--421}.

\bibitem[{Badier et~al.(1983)}]{Badier:1983mj}
\bibinfo{author}{J.~Badier}, et~al., \bibinfo{title}{{Experimental
  determination of the {$\pi$}-meson structure functions by the Drell-Yan
  mechanism}}, \bibinfo{journal}{Z. Phys. C} \bibinfo{volume}{18}
  (\bibinfo{year}{1983}) \bibinfo{pages}{281}.

\bibitem[{Betev et~al.(1985)}]{Betev:1985pg}
\bibinfo{author}{B.~Betev}, et~al., \bibinfo{title}{{Observation of anomalous
  scaling violation in muon pair production by 194-GeV/c {$\pi$}-tungsten
  interactions}}, \bibinfo{journal}{Z. Phys. C} \bibinfo{volume}{28}
  (\bibinfo{year}{1985}) \bibinfo{pages}{15}.

\bibitem[{Conway et~al.(1989)}]{Conway:1989fs}
\bibinfo{author}{J.~S. Conway}, et~al., \bibinfo{title}{{Experimental study of
  muon pairs produced by 252-GeV pions on tungsten}}, \bibinfo{journal}{Phys.
  Rev. D} \bibinfo{volume}{39} (\bibinfo{year}{1989}) \bibinfo{pages}{92--122}.

\bibitem[{Amendolia et~al.(1984)}]{Amendolia:1984nz}
\bibinfo{author}{S.~R. Amendolia}, et~al., \bibinfo{title}{{A Measurement of
  the Pion Charge Radius}}, \bibinfo{journal}{Phys. Lett. B}
  \bibinfo{volume}{146} (\bibinfo{year}{1984}) \bibinfo{pages}{116}.

\bibitem[{Amendolia et~al.(1986{\natexlab{a}})}]{Amendolia:1986wj}
\bibinfo{author}{S.~R. Amendolia}, et~al., \bibinfo{title}{{A Measurement of
  the Space - Like Pion Electromagnetic Form-Factor}}, \bibinfo{journal}{Nucl.
  Phys. B} \bibinfo{volume}{277} (\bibinfo{year}{1986}{\natexlab{a}})
  \bibinfo{pages}{168}.

\bibitem[{Volmer et~al.(2001)}]{Volmer:2000ek}
\bibinfo{author}{J.~Volmer}, et~al., \bibinfo{title}{{Measurement of the
  Charged Pion Electromagnetic Form-Factor}}, \bibinfo{journal}{Phys. Rev.
  Lett.} \bibinfo{volume}{86} (\bibinfo{year}{2001})
  \bibinfo{pages}{1713--1716}.

\bibitem[{Horn et~al.(2006)}]{Horn:2006tm}
\bibinfo{author}{T.~Horn}, et~al., \bibinfo{title}{{Determination of the
  Charged Pion Form Factor at {$Q^2=1.60$} and {$2.45 \,({\rm GeV/c})^2$}}},
  \bibinfo{journal}{Phys. Rev. Lett.} \bibinfo{volume}{97}
  (\bibinfo{year}{2006}) \bibinfo{pages}{192001}.

\bibitem[{Tadevosyan et~al.(2007)}]{Tadevosyan:2007yd}
\bibinfo{author}{V.~Tadevosyan}, et~al., \bibinfo{title}{{Determination of the
  pion charge form factor for {$Q^2=0.60- 1.60\,{\rm GeV}^2$}}},
  \bibinfo{journal}{Phys. Rev. C} \bibinfo{volume}{75} (\bibinfo{year}{2007})
  \bibinfo{pages}{055205}.

\bibitem[{Blok et~al.(2008)}]{Blok:2008jy}
\bibinfo{author}{H.~P. Blok}, et~al., \bibinfo{title}{{Charged pion form factor
  between $Q^2$=0.60 and 2.45 GeV$^2$. I. Measurements of the cross section for
  the ${^1}$H($e,e'\pi^+$)$n$ reaction}}, \bibinfo{journal}{Phys. Rev. C}
  \bibinfo{volume}{78} (\bibinfo{year}{2008}) \bibinfo{pages}{045202}.

\bibitem[{Huber et~al.(2008)}]{Huber:2008id}
\bibinfo{author}{G.~Huber}, et~al., \bibinfo{title}{{Charged pion form-factor
  between {$Q^2 = 0.60\,$GeV$^2$} and {$2.45\,$GeV$^2$}. II. Determination of,
  and results for, the pion form-factor}}, \bibinfo{journal}{Phys. Rev. C}
  \bibinfo{volume}{78} (\bibinfo{year}{2008}) \bibinfo{pages}{045203}.

\bibitem[{Holt and Roberts(2010)}]{Holt:2010vj}
\bibinfo{author}{R.~J. Holt}, \bibinfo{author}{C.~D. Roberts},
  \bibinfo{title}{{Distribution Functions of the Nucleon and Pion in the
  Valence Region}}, \bibinfo{journal}{Rev. Mod. Phys.} \bibinfo{volume}{82}
  (\bibinfo{year}{2010}) \bibinfo{pages}{2991--3044}.

\bibitem[{Peng and Qiu(2016)}]{Peng:2016ebs}
\bibinfo{author}{J.-C. Peng}, \bibinfo{author}{J.-W. Qiu}, \bibinfo{title}{{The
  Drell-Yan Process}}, \bibinfo{journal}{The Universe}
  \bibinfo{volume}{4}~(\bibinfo{number}{3}) (\bibinfo{year}{2016})
  \bibinfo{pages}{34--44}.

\bibitem[{Dove et~al.(2021)}]{SeaQuest:2021zxb}
\bibinfo{author}{J.~Dove}, et~al., \bibinfo{title}{{The asymmetry of antimatter
  in the proton}}, \bibinfo{journal}{Nature}
  \bibinfo{volume}{590}~(\bibinfo{number}{7847}) (\bibinfo{year}{2021})
  \bibinfo{pages}{561--565}.

\bibitem[{Gao et~al.(2017)Gao, Chang, Liu, Roberts, and Tandy}]{Gao:2017mmp}
\bibinfo{author}{F.~Gao}, \bibinfo{author}{L.~Chang}, \bibinfo{author}{Y.-X.
  Liu}, \bibinfo{author}{C.~D. Roberts}, \bibinfo{author}{P.~C. Tandy},
  \bibinfo{title}{{Exposing strangeness: projections for kaon electromagnetic
  form factors}}, \bibinfo{journal}{Phys. Rev. D}
  \bibinfo{volume}{96}~(\bibinfo{number}{3}) (\bibinfo{year}{2017})
  \bibinfo{pages}{034024}.

\bibitem[{Chen et~al.(2018)Chen, Ding, Chang, and Roberts}]{Chen:2018rwz}
\bibinfo{author}{M.~Chen}, \bibinfo{author}{M.~Ding},
  \bibinfo{author}{L.~Chang}, \bibinfo{author}{C.~D. Roberts},
  \bibinfo{title}{{Mass-dependence of pseudoscalar meson elastic form
  factors}}, \bibinfo{journal}{Phys. Rev. D} \bibinfo{volume}{98}
  (\bibinfo{year}{2018}) \bibinfo{pages}{091505(R)}.

\bibitem[{Ding et~al.(2020{\natexlab{a}})Ding, Raya, Binosi, Chang, Roberts,
  and Schmidt}]{Ding:2019qlr}
\bibinfo{author}{M.~Ding}, \bibinfo{author}{K.~Raya},
  \bibinfo{author}{D.~Binosi}, \bibinfo{author}{L.~Chang},
  \bibinfo{author}{C.~D. Roberts}, \bibinfo{author}{S.~M. Schmidt},
  \bibinfo{title}{{Drawing insights from pion parton distributions}},
  \bibinfo{journal}{Chin. Phys. C (Lett.)} \bibinfo{volume}{44}
  (\bibinfo{year}{2020}{\natexlab{a}}) \bibinfo{pages}{031002}.

\bibitem[{Ding et~al.(2020{\natexlab{b}})Ding, Raya, Binosi, Chang, Roberts,
  and Schmidt}]{Ding:2019lwe}
\bibinfo{author}{M.~Ding}, \bibinfo{author}{K.~Raya},
  \bibinfo{author}{D.~Binosi}, \bibinfo{author}{L.~Chang},
  \bibinfo{author}{C.~D. Roberts}, \bibinfo{author}{S.~M. Schmidt},
  \bibinfo{title}{{Symmetry, symmetry breaking, and pion parton
  distributions}}, \bibinfo{journal}{Phys. Rev. D}
  \bibinfo{volume}{101}~(\bibinfo{number}{5})
  (\bibinfo{year}{2020}{\natexlab{b}}) \bibinfo{pages}{054014}.

\bibitem[{Cui et~al.(2020{\natexlab{a}})Cui, Ding, Gao, Raya, Binosi, Chang,
  Roberts, Rodr\'{\i}guez-Quintero, and Schmidt}]{Cui:2020tdf}
\bibinfo{author}{Z.-F. Cui}, \bibinfo{author}{M.~Ding},
  \bibinfo{author}{F.~Gao}, \bibinfo{author}{K.~Raya},
  \bibinfo{author}{D.~Binosi}, \bibinfo{author}{L.~Chang},
  \bibinfo{author}{C.~D. Roberts},
  \bibinfo{author}{J.~Rodr\'{\i}guez-Quintero}, \bibinfo{author}{S.~M.
  Schmidt}, \bibinfo{title}{{Kaon and pion parton distributions}},
  \bibinfo{journal}{Eur. Phys. J. C} \bibinfo{volume}{80}
  (\bibinfo{year}{2020}{\natexlab{a}}) \bibinfo{pages}{1064}.

\bibitem[{Dokshitzer(1977)}]{Dokshitzer:1977sg}
\bibinfo{author}{Y.~L. Dokshitzer}, \bibinfo{title}{Calculation of the
  Structure Functions for Deep Inelastic Scattering and $e^+$ $e^-$
  Annihilation by Perturbation Theory in Quantum Chromodynamics. ({\mbox {I}n
  {R}ussian})}, \bibinfo{journal}{Sov. Phys. JETP} \bibinfo{volume}{46}
  (\bibinfo{year}{1977}) \bibinfo{pages}{641--653}.

\bibitem[{Gribov and Lipatov(1971)}]{Gribov:1971zn}
\bibinfo{author}{V.~N. Gribov}, \bibinfo{author}{L.~N. Lipatov},
  \bibinfo{title}{{Deep inelastic electron scattering in perturbation theory}},
  \bibinfo{journal}{Phys. Lett. B} \bibinfo{volume}{37} (\bibinfo{year}{1971})
  \bibinfo{pages}{78--80}.

\bibitem[{Lipatov(1975)}]{Lipatov:1974qm}
\bibinfo{author}{L.~N. Lipatov}, \bibinfo{title}{{The parton model and
  perturbation theory}}, \bibinfo{journal}{Sov. J. Nucl. Phys.}
  \bibinfo{volume}{20} (\bibinfo{year}{1975}) \bibinfo{pages}{94--102}.

\bibitem[{Altarelli and Parisi(1977)}]{Altarelli:1977zs}
\bibinfo{author}{G.~Altarelli}, \bibinfo{author}{G.~Parisi},
  \bibinfo{title}{{Asymptotic Freedom in Parton Language}},
  \bibinfo{journal}{Nucl. Phys. B} \bibinfo{volume}{126} (\bibinfo{year}{1977})
  \bibinfo{pages}{298--318}.

\bibitem[{Cui et~al.(2020{\natexlab{b}})Cui, Zhang, Binosi, de~Soto, Mezrag,
  Papavassiliou, Roberts, Rodr{\'{\i}}guez-Quintero, Segovia, and
  Zafeiropoulos}]{Cui:2019dwv}
\bibinfo{author}{Z.-F. Cui}, \bibinfo{author}{J.-L. Zhang},
  \bibinfo{author}{D.~Binosi}, \bibinfo{author}{F.~de~Soto},
  \bibinfo{author}{C.~Mezrag}, \bibinfo{author}{J.~Papavassiliou},
  \bibinfo{author}{C.~D. Roberts},
  \bibinfo{author}{J.~Rodr{\'{\i}}guez-Quintero}, \bibinfo{author}{J.~Segovia},
  \bibinfo{author}{S.~Zafeiropoulos}, \bibinfo{title}{{Effective charge from
  lattice QCD}}, \bibinfo{journal}{Chin. Phys. C} \bibinfo{volume}{44}
  (\bibinfo{year}{2020}{\natexlab{b}}) \bibinfo{pages}{083102}.

\bibitem[{Cui et~al.(2022{\natexlab{a}})Cui, Ding, Morgado, Raya, Binosi,
  Chang, Papavassiliou, Roberts, Rodr\'\i{}guez-Quintero, and
  Schmidt}]{Cui:2021mom}
\bibinfo{author}{Z.~F. Cui}, \bibinfo{author}{M.~Ding}, \bibinfo{author}{J.~M.
  Morgado}, \bibinfo{author}{K.~Raya}, \bibinfo{author}{D.~Binosi},
  \bibinfo{author}{L.~Chang}, \bibinfo{author}{J.~Papavassiliou},
  \bibinfo{author}{C.~D. Roberts},
  \bibinfo{author}{J.~Rodr\'\i{}guez-Quintero}, \bibinfo{author}{S.~M.
  Schmidt}, \bibinfo{title}{{Concerning pion parton distributions}},
  \bibinfo{journal}{Eur. Phys. J. A} \bibinfo{volume}{58}~(\bibinfo{number}{1})
  (\bibinfo{year}{2022}{\natexlab{a}}) \bibinfo{pages}{10}.

\bibitem[{Cui et~al.(2022{\natexlab{b}})Cui, Ding, Morgado, Raya, Binosi,
  Chang, De~Soto, Roberts, Rodr\'\i{}guez-Quintero, and Schmidt}]{Cui:2022bxn}
\bibinfo{author}{Z.~F. Cui}, \bibinfo{author}{M.~Ding}, \bibinfo{author}{J.~M.
  Morgado}, \bibinfo{author}{K.~Raya}, \bibinfo{author}{D.~Binosi},
  \bibinfo{author}{L.~Chang}, \bibinfo{author}{F.~De~Soto},
  \bibinfo{author}{C.~D. Roberts},
  \bibinfo{author}{J.~Rodr\'\i{}guez-Quintero}, \bibinfo{author}{S.~M.
  Schmidt}, \bibinfo{title}{{Emergence of pion parton distributions}},
  \bibinfo{journal}{Phys. Rev. D} \bibinfo{volume}{105}~(\bibinfo{number}{9})
  (\bibinfo{year}{2022}{\natexlab{b}}) \bibinfo{pages}{L091502}.

\bibitem[{Grunberg(1980)}]{Grunberg:1980ja}
\bibinfo{author}{G.~Grunberg}, \bibinfo{title}{{Renormalization Group Improved
  Perturbative QCD}}, \bibinfo{journal}{Phys. Lett. B} \bibinfo{volume}{95}
  (\bibinfo{year}{1980}) \bibinfo{pages}{70}, \bibinfo{note}{[Erratum: Phys.
  Lett. B 110, 501 (1982)]}.

\bibitem[{Grunberg(1984)}]{Grunberg:1982fw}
\bibinfo{author}{G.~Grunberg}, \bibinfo{title}{{Renormalization Scheme
  Independent QCD and QED: The Method of Effective Charges}},
  \bibinfo{journal}{Phys. Rev. D} \bibinfo{volume}{29} (\bibinfo{year}{1984})
  \bibinfo{pages}{2315}.

\bibitem[{Dokshitzer(1998)}]{Dokshitzer:1998nz}
\bibinfo{author}{Y.~L. Dokshitzer}, \bibinfo{title}{{\emph{Perturbative QCD
  theory (includes our knowledge of \mbox{$\alpha(s)$})} - hep-ph/9812252}},
  in: \bibinfo{booktitle}{{High-energy physics. Proceedings, 29th International
  Conference, ICHEP'98, Vancouver, Canada, July 23-29, 1998. Vol. 1, 2}},
  \bibinfo{pages}{305--324}, \bibinfo{year}{1998}.

\bibitem[{Raya et~al.(2022)Raya, Cui, Chang, Morgado, Roberts, and
  Rodr{\'{\i}}guez-Quintero}]{Raya:2021zrz}
\bibinfo{author}{K.~Raya}, \bibinfo{author}{Z.-F. Cui},
  \bibinfo{author}{L.~Chang}, \bibinfo{author}{J.-M. Morgado},
  \bibinfo{author}{C.~D. Roberts},
  \bibinfo{author}{J.~Rodr{\'{\i}}guez-Quintero}, \bibinfo{title}{{Revealing
  pion and kaon structure via generalised parton distributions}},
  \bibinfo{journal}{Chin. Phys. C} \bibinfo{volume}{46}~(\bibinfo{number}{26})
  (\bibinfo{year}{2022}) \bibinfo{pages}{013105}.

\bibitem[{Deur et~al.(2016)Deur, Brodsky, and de~Teramond}]{Deur:2016tte}
\bibinfo{author}{A.~Deur}, \bibinfo{author}{S.~J. Brodsky},
  \bibinfo{author}{G.~F. de~Teramond}, \bibinfo{title}{{The QCD Running
  Coupling}}, \bibinfo{journal}{Prog. Part. Nucl. Phys.} \bibinfo{volume}{90}
  (\bibinfo{year}{2016}) \bibinfo{pages}{1--74}.

\bibitem[{Deur et~al.(2022)Deur, Burkert, Chen, and Korsch}]{Deur:2022msf}
\bibinfo{author}{A.~Deur}, \bibinfo{author}{V.~Burkert}, \bibinfo{author}{J.~P.
  Chen}, \bibinfo{author}{W.~Korsch}, \bibinfo{title}{{Experimental
  determination of the QCD effective charge $\alpha_{g_1}(Q)$}},
  \bibinfo{journal}{Particles} \bibinfo{volume}{5}~(\bibinfo{number}{2})
  (\bibinfo{year}{2022}) \bibinfo{pages}{171--179}.

\bibitem[{Deur et~al.(????)Deur, Brodsky, and Roberts}]{Deur:2023}
\bibinfo{author}{A.~Deur}, \bibinfo{author}{S.~J. Brodsky},
  \bibinfo{author}{C.~D. Roberts}, \bibinfo{title}{{QCD Running Couplings and
  Effective Charges -- arXiv:2303.00723 [hep-ph]}} .

\bibitem[{Aicher et~al.(2010)Aicher, Sch{\"a}fer, and
  Vogelsang}]{Aicher:2010cb}
\bibinfo{author}{M.~Aicher}, \bibinfo{author}{A.~Sch{\"a}fer},
  \bibinfo{author}{W.~Vogelsang}, \bibinfo{title}{{Soft-Gluon Resummation and
  the Valence Parton Distribution Function of the Pion}},
  \bibinfo{journal}{Phys.\ Rev.\ Lett.} \bibinfo{volume}{105}
  (\bibinfo{year}{2010}) \bibinfo{pages}{252003}.

\bibitem[{Barry et~al.(2021)Barry, Ji, Sato, and Melnitchouk}]{Barry:2021osv}
\bibinfo{author}{P.~C. Barry}, \bibinfo{author}{C.-R. Ji},
  \bibinfo{author}{N.~Sato}, \bibinfo{author}{W.~Melnitchouk},
  \bibinfo{title}{{Global QCD Analysis of Pion Parton Distributions with
  Threshold Resummation}}, \bibinfo{journal}{Phys. Rev. Lett.}
  \bibinfo{volume}{127}~(\bibinfo{number}{23}) (\bibinfo{year}{2021})
  \bibinfo{pages}{232001}.

\bibitem[{Jo\'o et~al.(2019)Jo\'o, Karpie, Orginos, Radyushkin, Richards,
  Sufian, and Zafeiropoulos}]{Joo:2019bzr}
\bibinfo{author}{B.~Jo\'o}, \bibinfo{author}{J.~Karpie},
  \bibinfo{author}{K.~Orginos}, \bibinfo{author}{A.~V. Radyushkin},
  \bibinfo{author}{D.~G. Richards}, \bibinfo{author}{R.~S. Sufian},
  \bibinfo{author}{S.~Zafeiropoulos}, \bibinfo{title}{{Pion valence structure
  from Ioffe-time parton pseudodistribution functions}},
  \bibinfo{journal}{Phys. Rev. D} \bibinfo{volume}{100} (\bibinfo{year}{2019})
  \bibinfo{pages}{114512}.

\bibitem[{Sufian et~al.(2019)Sufian, Karpie, Egerer, Orginos, Qiu, and
  Richards}]{Sufian:2019bol}
\bibinfo{author}{R.~S. Sufian}, \bibinfo{author}{J.~Karpie},
  \bibinfo{author}{C.~Egerer}, \bibinfo{author}{K.~Orginos},
  \bibinfo{author}{J.-W. Qiu}, \bibinfo{author}{D.~G. Richards},
  \bibinfo{title}{{Pion Valence Quark Distribution from Matrix Element
  Calculated in Lattice QCD}}, \bibinfo{journal}{Phys. Rev. D}
  \bibinfo{volume}{99} (\bibinfo{year}{2019}) \bibinfo{pages}{074507}.

\bibitem[{Alexandrou et~al.(2021)Alexandrou, Bacchio, Cloet, Constantinou,
  Hadjiyiannakou, Koutsou, and Lauer}]{Alexandrou:2021mmi}
\bibinfo{author}{C.~Alexandrou}, \bibinfo{author}{S.~Bacchio},
  \bibinfo{author}{I.~Cloet}, \bibinfo{author}{M.~Constantinou},
  \bibinfo{author}{K.~Hadjiyiannakou}, \bibinfo{author}{G.~Koutsou},
  \bibinfo{author}{C.~Lauer}, \bibinfo{title}{{Pion and kaon $\langle
  x^3\rangle$ from lattice QCD and PDF reconstruction from Mellin moments}},
  \bibinfo{journal}{Phys. Rev. D} \bibinfo{volume}{104}~(\bibinfo{number}{5})
  (\bibinfo{year}{2021}) \bibinfo{pages}{054504}.

\bibitem[{Belitsky and Radyushkin(2005)}]{Belitsky:2005qn}
\bibinfo{author}{A.~Belitsky}, \bibinfo{author}{A.~Radyushkin},
  \bibinfo{title}{{Unraveling hadron structure with generalized parton
  distributions}}, \bibinfo{journal}{Phys. Rept.} \bibinfo{volume}{418}
  (\bibinfo{year}{2005}) \bibinfo{pages}{1--387}.

\bibitem[{Mezrag(2022)}]{Mezrag:2022pqk}
\bibinfo{author}{C.~Mezrag}, \bibinfo{title}{{An Introductory Lecture on
  Generalised Parton Distributions}}, \bibinfo{journal}{Few Body Syst.}
  \bibinfo{volume}{63}~(\bibinfo{number}{3}) (\bibinfo{year}{2022})
  \bibinfo{pages}{62}.

\bibitem[{Mezrag(2023)}]{Mezrag:2023}
\bibinfo{author}{C.~Mezrag}, \bibinfo{title}{{Generalised Parton Distributions
  in Continuum Schwinger Methods: progress, opportunities and challenges}},
  \bibinfo{journal}{Particles} \bibinfo{volume}{6}~(\bibinfo{number}{1})
  (\bibinfo{year}{2023}) \bibinfo{pages}{262--296}.

\bibitem[{Xu et~al.(2018)Xu, Chang, Roberts, and Zong}]{Xu:2018eii}
\bibinfo{author}{S.-S. Xu}, \bibinfo{author}{L.~Chang}, \bibinfo{author}{C.~D.
  Roberts}, \bibinfo{author}{H.-S. Zong}, \bibinfo{title}{{Pion and kaon
  valence-quark parton quasidistributions}}, \bibinfo{journal}{Phys. Rev. D}
  \bibinfo{volume}{97} (\bibinfo{year}{2018}) \bibinfo{pages}{094014}.

\bibitem[{Zhang et~al.(2021)Zhang, Raya, Chang, Cui, Morgado, Roberts, and
  Rodr\'\i{}guez-Quintero}]{Zhang:2021mtn}
\bibinfo{author}{J.-L. Zhang}, \bibinfo{author}{K.~Raya},
  \bibinfo{author}{L.~Chang}, \bibinfo{author}{Z.-F. Cui},
  \bibinfo{author}{J.~M. Morgado}, \bibinfo{author}{C.~D. Roberts},
  \bibinfo{author}{J.~Rodr\'\i{}guez-Quintero}, \bibinfo{title}{{Measures of
  pion and kaon structure from generalised parton distributions}},
  \bibinfo{journal}{Phys. Lett. B} \bibinfo{volume}{815} (\bibinfo{year}{2021})
  \bibinfo{pages}{136158}.

\bibitem[{Cui et~al.(2021)Cui, Binosi, Roberts, and Schmidt}]{Cui:2021aee}
\bibinfo{author}{Z.-F. Cui}, \bibinfo{author}{D.~Binosi},
  \bibinfo{author}{C.~D. Roberts}, \bibinfo{author}{S.~M. Schmidt},
  \bibinfo{title}{{Pion charge radius from pion+electron elastic scattering
  data}}, \bibinfo{journal}{Phys. Lett. B} \bibinfo{volume}{822}
  (\bibinfo{year}{2021}) \bibinfo{pages}{136631}.

\bibitem[{Lepage and Brodsky(1979)}]{Lepage:1979za}
\bibinfo{author}{G.~Lepage}, \bibinfo{author}{S.~J. Brodsky},
  \bibinfo{title}{{Exclusive Processes in Quantum Chromodynamics: The
  Form-Factors of Baryons at Large Momentum Transfer}}, \bibinfo{journal}{Phys.
  Rev. Lett.} \bibinfo{volume}{43} (\bibinfo{year}{1979})
  \bibinfo{pages}{545--549}, \bibinfo{note}{[Erratum: Phys. Rev. Lett. 43,
  1625--1626 (1979)]}.

\bibitem[{Efremov and Radyushkin(1980)}]{Efremov:1979qk}
\bibinfo{author}{A.~V. Efremov}, \bibinfo{author}{A.~V. Radyushkin},
  \bibinfo{title}{{Factorization and Asymptotical Behavior of Pion Form- Factor
  in QCD}}, \bibinfo{journal}{Phys. Lett. B} \bibinfo{volume}{94}
  (\bibinfo{year}{1980}) \bibinfo{pages}{245--250}.

\bibitem[{Lepage and Brodsky(1980)}]{Lepage:1980fj}
\bibinfo{author}{G.~P. Lepage}, \bibinfo{author}{S.~J. Brodsky},
  \bibinfo{title}{{Exclusive Processes in Perturbative Quantum
  Chromodynamics}}, \bibinfo{journal}{Phys. Rev. D} \bibinfo{volume}{22}
  (\bibinfo{year}{1980}) \bibinfo{pages}{2157--2198}.

\bibitem[{Polyakov and Schweitzer(2018)}]{Polyakov:2018zvc}
\bibinfo{author}{M.~V. Polyakov}, \bibinfo{author}{P.~Schweitzer},
  \bibinfo{title}{{Forces inside hadrons: pressure, surface tension, mechanical
  radius, and all that}}, \bibinfo{journal}{Int. J. Mod. Phys. A}
  \bibinfo{volume}{33}~(\bibinfo{number}{26}) (\bibinfo{year}{2018})
  \bibinfo{pages}{1830025}.

\bibitem[{Masuda et~al.(2016)}]{Belle:2015oin}
\bibinfo{author}{M.~Masuda}, et~al., \bibinfo{title}{{Study of $\pi^0$ pair
  production in single-tag two-photon collisions}}, \bibinfo{journal}{Phys.
  Rev. D} \bibinfo{volume}{93}~(\bibinfo{number}{3}) (\bibinfo{year}{2016})
  \bibinfo{pages}{032003}.

\bibitem[{Kumano et~al.(2018)Kumano, Song, and Teryaev}]{Kumano:2017lhr}
\bibinfo{author}{S.~Kumano}, \bibinfo{author}{Q.-T. Song},
  \bibinfo{author}{O.~V. Teryaev}, \bibinfo{title}{{Hadron tomography by
  generalized distribution amplitudes in pion-pair production process $\gamma^*
  \gamma \rightarrow \pi^0 \pi^0 $ and gravitational form factors for pion}},
  \bibinfo{journal}{Phys. Rev. D} \bibinfo{volume}{97} (\bibinfo{year}{2018})
  \bibinfo{pages}{014020}.

\bibitem[{Shastry et~al.(2022)Shastry, Broniowski, and
  Ruiz~Arriola}]{Shastry:2022obb}
\bibinfo{author}{V.~Shastry}, \bibinfo{author}{W.~Broniowski},
  \bibinfo{author}{E.~Ruiz~Arriola}, \bibinfo{title}{{Generalized quasi-,
  Ioffe-time-, and pseudodistributions of the pion in the
  Nambu\textendash{}Jona-Lasinio model}}, \bibinfo{journal}{Phys. Rev. D}
  \bibinfo{volume}{106}~(\bibinfo{number}{11}) (\bibinfo{year}{2022})
  \bibinfo{pages}{114035}.

\bibitem[{Xing et~al.(2023)Xing, Ding, and Chang}]{Xing:2022mvk}
\bibinfo{author}{Z.~Xing}, \bibinfo{author}{M.~Ding},
  \bibinfo{author}{L.~Chang}, \bibinfo{title}{{Glimpse into the pion
  gravitational form factor}}, \bibinfo{journal}{Phys. Rev. D}
  \bibinfo{volume}{107}~(\bibinfo{number}{3}) (\bibinfo{year}{2023})
  \bibinfo{pages}{L031502}.

\bibitem[{Chen et~al.(2013)Chen, Chang, Roberts, Wan, Schmidt, and
  Wilson}]{Chen:2012txa}
\bibinfo{author}{C.~Chen}, \bibinfo{author}{L.~Chang}, \bibinfo{author}{C.~D.
  Roberts}, \bibinfo{author}{S.-L. Wan}, \bibinfo{author}{S.~M. Schmidt},
  \bibinfo{author}{D.~J. Wilson}, \bibinfo{title}{{Features and flaws of a
  contact interaction treatment of the kaon}}, \bibinfo{journal}{Phys. Rev. C}
  \bibinfo{volume}{87} (\bibinfo{year}{2013}) \bibinfo{pages}{045207}.

\bibitem[{Qin et~al.(2018)Qin, Chen, Mezrag, and Roberts}]{Qin:2017lcd}
\bibinfo{author}{S.-X. Qin}, \bibinfo{author}{C.~Chen},
  \bibinfo{author}{C.~Mezrag}, \bibinfo{author}{C.~D. Roberts},
  \bibinfo{title}{{Off-shell persistence of composite pions and kaons}},
  \bibinfo{journal}{Phys. Rev. C} \bibinfo{volume}{97} (\bibinfo{year}{2018})
  \bibinfo{pages}{015203}.

\bibitem[{Dally et~al.(1980)Dally, Hauptman, Kubic, Stork, Watson
  et~al.}]{Dally:1980dj}
\bibinfo{author}{E.~Dally}, \bibinfo{author}{J.~Hauptman},
  \bibinfo{author}{J.~Kubic}, \bibinfo{author}{D.~Stork},
  \bibinfo{author}{A.~Watson}, et~al., \bibinfo{title}{{Direct Measurement of
  the Negative-Kaon Form Factor}}, \bibinfo{journal}{Phys. Rev. Lett.}
  \bibinfo{volume}{45} (\bibinfo{year}{1980}) \bibinfo{pages}{232--235}.

\bibitem[{Amendolia et~al.(1986{\natexlab{b}})}]{Amendolia:1986ui}
\bibinfo{author}{S.~Amendolia}, et~al., \bibinfo{title}{{A Measurement of the
  Kaon Charge Radius}}, \bibinfo{journal}{Phys. Lett. B} \bibinfo{volume}{178}
  (\bibinfo{year}{1986}{\natexlab{b}}) \bibinfo{pages}{435}.

\bibitem[{Badier et~al.(1980)}]{Badier:1980jq}
\bibinfo{author}{J.~Badier}, et~al., \bibinfo{title}{{Measurement of the {$K^-
  / \pi^-$} structure function ratio using the Drell-Yan process}},
  \bibinfo{journal}{Phys. Lett. B} \bibinfo{volume}{93} (\bibinfo{year}{1980})
  \bibinfo{pages}{354}.

\bibitem[{Punjabi et~al.(2015)Punjabi, Perdrisat, Jones, Brash, and
  Carlson}]{Punjabi:2015bba}
\bibinfo{author}{V.~Punjabi}, \bibinfo{author}{C.~F. Perdrisat},
  \bibinfo{author}{M.~K. Jones}, \bibinfo{author}{E.~J. Brash},
  \bibinfo{author}{C.~E. Carlson}, \bibinfo{title}{{The Structure of the
  Nucleon: Elastic Electromagnetic Form Factors}}, \bibinfo{journal}{Eur. Phys.
  J. A} \bibinfo{volume}{51} (\bibinfo{year}{2015}) \bibinfo{pages}{79}.

\bibitem[{Gao and Vanderhaeghen(2022)}]{Gao:2021sml}
\bibinfo{author}{H.~Gao}, \bibinfo{author}{M.~Vanderhaeghen},
  \bibinfo{title}{{The proton charge radius}}, \bibinfo{journal}{Rev. Mod.
  Phys.} \bibinfo{volume}{94}~(\bibinfo{number}{1}) (\bibinfo{year}{2022})
  \bibinfo{pages}{015002}.

\bibitem[{Cui et~al.(2022{\natexlab{c}})Cui, Binosi, Roberts, and
  Schmidt}]{Cui:2022fyr}
\bibinfo{author}{Z.-F. Cui}, \bibinfo{author}{D.~Binosi},
  \bibinfo{author}{C.~D. Roberts}, \bibinfo{author}{S.~M. Schmidt},
  \bibinfo{title}{{Hadron and light nucleus radii from electron scattering}},
  \bibinfo{journal}{Chin. Phys. C} \bibinfo{volume}{46}~(\bibinfo{number}{12})
  (\bibinfo{year}{2022}{\natexlab{c}}) \bibinfo{pages}{122001}.

\bibitem[{Lu et~al.(2022)Lu, Chang, Raya, Roberts, and
  Rodr\'\i{}guez-Quintero}]{Lu:2022cjx}
\bibinfo{author}{Y.~Lu}, \bibinfo{author}{L.~Chang}, \bibinfo{author}{K.~Raya},
  \bibinfo{author}{C.~D. Roberts},
  \bibinfo{author}{J.~Rodr\'\i{}guez-Quintero}, \bibinfo{title}{{Proton and
  pion distribution functions in counterpoint}}, \bibinfo{journal}{Phys. Lett.
  B} \bibinfo{volume}{830} (\bibinfo{year}{2022}) \bibinfo{pages}{137130}.

\end{thebibliography}

\end{document}